%% file: mgirardiA523REV.tex
\PassOptionsToPackage{pdfpagelabels=false}{hyperref}
\documentclass[fleqn,usenatbib]{mnras}

\usepackage{mathptmx}
\usepackage[T1]{fontenc}
\usepackage{ae,aecompl}

\usepackage{graphicx}	% Including figure files
\newcommand{\mincir}{\raise -2.truept\hbox{\rlap{\hbox{$\sim$}}\raise5.truept
\hbox{$<$}\ }}
\newcommand{\magcir}{\raise -2.truept\hbox{\rlap{\hbox{$\sim$}}\raise5.truept
\hbox{$>$}\ }}
\newcommand{\siml}{\raise -2.truept\hbox{\rlap{\hbox{$\sim$}}\raise5.truept
\hbox{$<$}\ }}
\newcommand{\simg}{\raise -2.truept\hbox{\rlap{\hbox{$\sim$}}\raise5.truept
\hbox{$>$}\ }}
\newcommand{\be}{\begin{equation}}
\newcommand{\ee}{\end{equation}}
\newcommand{\ba}{\begin{eqnarray}}
\newcommand{\ea}{\end{eqnarray}}
\newcommand {\kpc} {$h_{70}^{-1}$ kpc $\;$}
\newcommand {\kpcc} {$h_{70}^{-1}$ kpc}

\newcommand {\h} {$h_{70}^{-1}$ Mpc$\;$}
\newcommand {\hh} {$h_{70}^{-1}$ Mpc}
\newcommand {\hhh} {\;h_{70}^{-1} \mathrm{Mpc}}
\newcommand {\ks} {km~s$^{-1} \;$}
\newcommand {\kss} {km~s$^{-1}$}

\newcommand {\mqua} {$\times 10^{14}\;h_{70}^{-1}\;M_{\odot} \;$}
\newcommand {\mquaa} {$\times 10^{14}\;h_{70}^{-1}\;M_{\odot}$}

\newcommand{\fxunits}{\mbox{erg cm$^{-2}$ s$^{-1}$}}
\newcommand {\lxunits} {$h_{70}^{-2}$ erg\ s$^{-1}$}
\newcommand{\degree}{\ensuremath{\mathrm{^\circ}}}
\newcommand{\arcm}{\ensuremath{\mathrm{^\prime}\;}}
\newcommand{\arcs}{\ensuremath{\arcmm\hskip -0.1em\arcmm \;}}
\newcommand{\arcmm}{\ensuremath{\mathrm{^\prime}}}
\newcommand{\arcss}{\ensuremath{\arcmm\hskip -0.1em\arcmm}}
\newcommand{\dotarcs}{\,\rlap{\hbox{$\mathrm{^\prime\hskip-0.1em^\prime}$}}{\hbox{$.$}}\,}

\newcommand{\dotsec}{\,\rlap{\hbox{$\mathrm{^s}$}}{\hbox{$.$}}\,}
\newcommand{\chandra }{{\em Chandra}}
\newcommand{\xspec }{{\em Xspec}}
\newcommand{\acisi }{{\em ACIS-I}}
\newcommand{\ciao }{{\em CIAO}}

\newcommand\rosat{{\sl ROSAT}}

\title[A multiwavelength view of Abell 523]{A multiwavelength view of the galaxy cluster Abell 523 and its peculiar
diffuse radio source}

\author[M. Girardi et al.]{M. Girardi,$^{1,2}$\thanks{E-mail: girardi@oats.inaf.it}
W. Boschin,$^{3,4,5}$
F. Gastaldello,$^{6,7}$
G. Giovannini,$^{8,9}$
F. Govoni,$^{10}$
\newauthor M. Murgia,$^{10}$
R. Barrena,$^{4,5}$
S. Ettori,$^{11,12}$
M. Trasatti,$^{8,9}$
V. Vacca$^{10,13}$ 
\\
$^{1}$Dipartimento di Fisica, Universit\`a degli Studi di Trieste, via
Tiepolo 11, I-34143 Trieste, Italy\\ 
$^{2}$INAF-Osservatorio
Astronomico di Trieste, via Tiepolo 11, I-34143 Trieste,
Italy\\ 
$^{3}$Fundaci\'on G. Galilei - INAF (Telescopio Nazionale
Galileo), Rambla J. A. Fern\'andez P\'erez 7, E-38712 Bre\~na Baja (La
Palma), Spain\\ 
$^{4}$Instituto de Astrof\'{\i}sica de Canarias,
C/V\'{\i}a L\'actea s/n, E-38205 La Laguna (Tenerife),
Spain\\ 
$^{5}$Departamento de Astrof\'{\i}sica, Univ. de La Laguna,
Av. del Astrof\'{\i}sico Francisco S\'anchez s/n, E-38205 La Laguna
(Tenerife), Spain\\ 
$^{6}$INAF-IASF Milano, Via Bassini 15, I-20133
Milano, Italy\\ 
$^{7}$Department of Physics and Astronomy, University
of California at Irvine, 4129 Frederick Reines Hall, Irvine, CA
92697-4575, USA\\
$^{8}$Dipartimento di Fisica e Astronomia,
Universit\`a degli Studi di Bologna, via Ranzani 1, I-40127 Bologna,
Italy\\ 
$^{9}$INAF-Osservatorio di Radioastronomia, via Gobetti 101,
I-40129 Bologna, Italy\\ 
$^{10}$INAF-Osservatorio Astronomico di
Cagliari, Via della Scienza 5, 09047 Selargius, Cagliari, Italy\\ 
$^{11}$INAF-Osservatorio Astronomico di Bologna, via
Ranzani 1, 40127, Bologna, Italy\\ 
$^{12}$INFN - Sezione di Bologna,
viale Berti Pichat 6/2, I-40127, Bologna, Italy\\ 
$^{13}$Max Planck Institute for Astrophysics, Karl-Schwarzschild-Str. 1, 85748 Garching, Germany\\}

\date{Accepted XXX. Received YYY; in original form ZZZ}
\pubyear{2015}

\begin{document}
\label{firstpage}
\pagerange{\pageref{firstpage}--\pageref{lastpage}}
\maketitle

\begin{abstract}

We study the structure of the galaxy cluster Abell 523 (A523) at
$z=0.104$ using new spectroscopic data for 132 galaxies acquired at
the Telescopio Nazionale {\em Galileo}, new photometric data from the
{\em Isaac Newton} Telescope, and X-ray and radio data from
the \chandra\ and Very Large Array archives. We estimate the velocity
dispersion of the galaxy population, $\sigma_{\rm
V}=949_{-60}^{+80}$ \kss, and the X-ray temperature of the hot
intracluster medium, $kT=5.3\pm0.3$ keV. We infer that A523 is a
massive system: $M_{200}\sim 7-9$ $\times 10^{14}\;M_{\odot}$. The
analysis of the optical data confirms the presence of two subclusters,
0.75 Mpc apart, tracing the SSW-NNE direction and dominated by the two
brightest cluster galaxies (BCG1 and BCG2). The X-ray surface
brightness is strongly elongated towards the NNE direction, and its
peak is clearly offset from both the BCGs.  We confirm the presence of
a 1.3 Mpc large radio halo, elongated in the ESE-WNW direction and
perpendicular to the optical/X-ray elongation.  We detect a
significant radio/X-ray offset and radio polarization, two features
which might be the result of a magnetic field energy spread on large
spatial scales.  A523 is found consistent with most scaling relations
followed by clusters hosting radio haloes, but quite peculiar in the
$P_{\rm radio}$-$L_{\rm X}$ relation: it is underluminous in the
X-rays or overluminous in radio. A523 can be described as a binary
head--on merger caught after a collision along the SSW-NNE direction.
However, minor optical and radio features suggest a more complex
cluster structure, with A523 forming at the crossing of two filaments
along the SSW-NNE and ESE-WNW directions.

\end{abstract}

\begin{keywords}
galaxies: clusters: general -- galaxies: clusters: individual: Abell
523 -- galaxies: kinematics and dynamics -- radio continuum: general--
X-rays: galaxies: clusters.
\end{keywords}

%
%________________________________________________________________

\section{INTRODUCTION}
\label{intro}

Clusters of galaxies are the largest gravitationally bound systems in
the Universe, with typical masses of about $10^{14}$-$10^{15}$
M$_{\sun}$.  Clusters are formed by hierarchical formation processes,
where smaller units (galaxies, groups and small clusters) formed first
and merged under gravitational pull to larger and larger units in the
course of time, clusters forming at the intersection of filaments of
the large scale structure of the Universe. Merging processes are the
subject of multiwavelength studies (e.g., Feretti et
al. \citeyear{fer02}). Major cluster mergers are among the most
energetic events in the Universe since the Big Bang (e.g., Sarazin
\citeyear{sar02}).

An increasing number of galaxy clusters shows the presence of diffuse
radio emission on Mpc scale (e.g., Feretti \citeyear{fer05}; Ferrari
\citeyear{fer08}; Venturi \citeyear{ven11}; Feretti et
al. \citeyear{fer12}).  The synchrotron emission of these radio sources
proves the existence of a large-scale magnetic field and relativistic
particles spread out in the cluster.  These extended diffuse radio
sources are usually classified as relics and haloes, having quite different
phenomenological properties (e.g., Giovannini \&
Feretti \citeyear{gio04}; Feretti et al. \citeyear{fer12}).  For both
types of sources, cluster mergers have been proposed as the process
responsible for their origin (e.g., Trimble \citeyear{tri93}; Feretti
\citeyear{fer99}; Brunetti \& Jones \citeyear{bru15}).  This scenario
is supported by an increasing amount of observational evidence (e.g.,
Schuecker et al. \citeyear{sch01}; Buote \citeyear{buo02}; Cassano et
al. \citeyear{cas10}; Rossetti et al. \citeyear{ros11}; Girardi et
al. \citeyear{gir11} and references therein).

Relics are elongated sources, found at the cluster outskirts, with the
major axis roughly perpendicular to the direction of the cluster
merger (e.g., van Weeren et al. \citeyear{vanwee11a}).  When observed
with high angular resolution, they generally show an asymmetric
transverse profile, with a sharp edge usually on the side towards the
cluster outer edge. These morphologies are in very good agreement with
models predicting that these sources are related to large-scale shocks
generated during cluster merger events (e.g., En{\ss}lin et
al. \citeyear{ens98}; Br\"uggen et al. \citeyear{bru11}). Instead,
radio haloes have rounder morphologies, are unpolarized and fill the
central cluster regions occupied by the X-ray emitting ICM. A close
similarity between the radio and the X-ray morphology has been found
in a number of clusters hosting a radio halo (e.g., Govoni et
al. \citeyear{gov01a}). Radio haloes are probably related to turbulent
motions of ICM due to merger events (e.g., Brunetti et
al. \citeyear{bru09}; Brunetti \& Jones \citeyear{bru15}).

There are also diffuse radio emissions with peculiar properties, e.g.,
relic sources with a roundish structure (Feretti et
al. \citeyear{fer06}), the relic of 1RXS J0603.3+4214 with an unusual
morphology (Ogrean et al. \citeyear{ogr13}), haloes having offsets
between the radio and the X-ray peak (Govoni et.
al. \citeyear{gov12}). In addition, examples of bridges between relics
and haloes have been observed in a few clusters (e.g., Coma cluster --
Kim et al. \citeyear{kim89}; Abell 2744 -- Govoni et
al. \citeyear{gov01a}), and diffuse radio sources at large distances
from a few clusters have been detected (e.g., Abell 2255 -- Pizzo et
al. \citeyear{piz08}; Abell 2256 -- van Weeren et
al. \citeyear{vanwee09}). An intriguing case is also that of the
diffuse source 0809+39 (Brown \& Rudnick
\citeyear{bro09}), where the southern component is possibly associated with an
$\sim$5 Mpc long galaxy filament at $z\sim$0.04.

The present paper is focused on the study of Abell 523 (hereafter
A523). A523 hosts an extended and powerful diffuse emission with
a maximum linear size of $\sim 1.3$ \h and a total radio power of $P_{\rm
  1.4\ GHz} \sim 1.5\ 10^{24}$ W/Hz, strongly elongated along the
ESE-WNW direction (Giovannini et al. \citeyear{gio11}, hereafter G11).  G11
classified this radio source as a radio halo because of (i) the
radio emission permeates both the merging clumps and (ii) the
elongated structure does not show any morphological feature typical of
radio relics such as high brightness filamentary structures or a
transverse flux asymmetry (see e.g., van Weeren et
al. \citeyear{vanwee11b}).

\begin{figure*}
\centering 
\includegraphics[width=18cm]{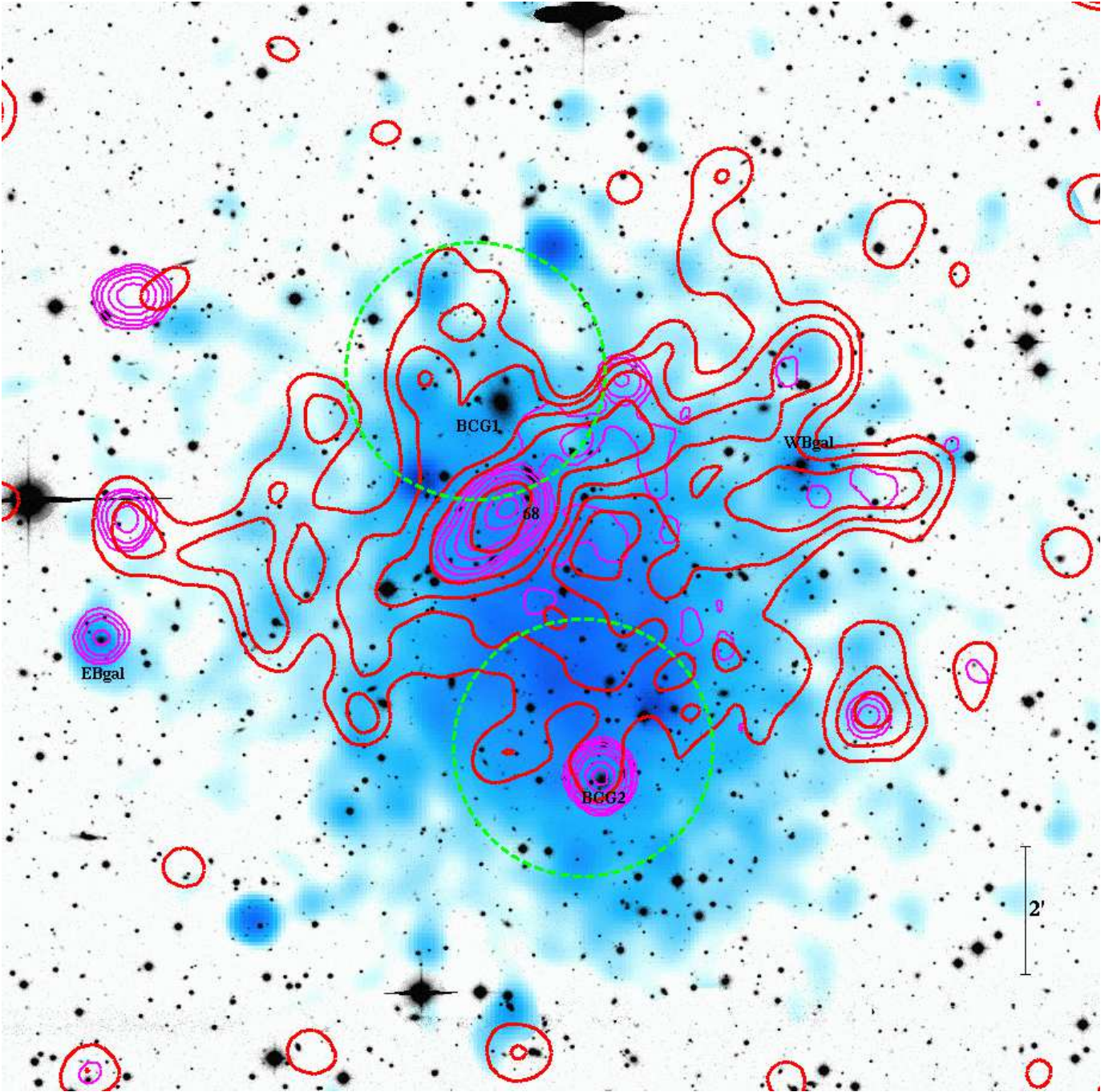}
\caption{Multiwavelength view of Abell 523. The grey-scale image in
background corresponds to the $r$-band (INT data). Superimposed, with
blue colour, we show the smoothed \chandra\ X-ray image in the 0.5-2
keV energy range (see Sect.~\ref{xray}).  Thin magenta contours refer
to the high--resolution 1.4 GHz VLA image and highlight the discrete
radio sources.  Thick red contours refer to the low-resolution 1.4 GHz
VLA image (with discrete sources subtracted) and highlight the radio
halo (see Sect.~\ref{radio}). Dashed green circles indicate the
centres of the two subclusters detected through the 2D-DEDICA analysis
applied to the photometric member galaxies (see
Sect.~\ref{data}). Labels refer to specific galaxies discussed in the
text.  North is up and east is left.  At the cluster redshift,
the scale of 2\arcm correspond to $\sim 0.23$ \h and the FOW
of the whole image is $2\times2$ \hh.  }
\label{figimage}
\end{figure*}

However, the A523 radio halo phenomenology is far from the typical
one. In fact, (i) the radio halo is strongly elongated in the
direction perpendicular to the likely merging axis (see fig.~3 of G11
and our Fig.~\ref{figimage}), while, generally, the opposite
phenomenology is observed, with the radio halo being elongated in the
same direction as the merger (e.g., Abell 520 - Govoni et
al. \citeyear{gov01b}; Girardi et al. \citeyear{gir08}; Abell 2255
 - Govoni et al. \citeyear{gov05}; Abell 665 - Giovannini \& Feretti
\citeyear{gio00}).  More quantitatively, (ii) A523 is peculiar in its
deviation from the typical $P_{\rm radio}-L_{\rm X}$ relation having a
higher radio power or a lower X-ray luminosity than expected given 
the estimate of the X-ray luminosity from \rosat\ data ($L_{\rm
  X,0.1-2.4\ keV}\sim 1\times 10^{44}$ erg s$^{-1}$, Ebeling et
al. \citeyear{ebe98}; B\"ohringer et al. \citeyear{boe00}).

On the basis of data available at that time, G11 analysed the internal
structure of A523.  The analysis of a colour selected photometric
catalogues of galaxies extracted from the SuperCOSMOS Sky Surveys 
showed that the cluster is bimodal and elongated along the SSW-NNE
direction. The analysis of \rosat\ X-ray data revealed a bimodal
distribution of the hot ICM,  with a clear shift between
the galaxies and gas distribution, thus suggesting an ongoing
merger (see fig.~5 of G11).  Unfortunately, so far, only the redshift
of the brightest cluster galaxy (hereafter BCG1) has been available and the
kinematical analysis based on member galaxies has never been performed.

A523 was included in our observational program aimed to study the
internal dynamics of clusters with diffuse radio emission (DARC --
Dynamical Analysis of Radio Clusters, see Girardi et
al. \citeyear{gir11} and refs. therein\footnote{see also
http://adlibitum.oat.ts.astro.it/girardi/darc, the web site of the
DARC project.}). We obtained new photometric and spectroscopic data
acquired at the {\em Isaac Newton} Telescope (INT) and at the
Telescopio Nazionale {\em Galileo} (TNG). We also analysed very
recent, never analysed before, X-ray data from the \chandra\ archive and
radio data from the Jansky Very Large Array (VLA) archive data.

This paper is organized as follows. We describe optical observations
and present our spectroscopic data catalogue in Section~2, while
Section~3 presents our analysis and results based on optical galaxy
data. In Sections~4 and 5 we present our analyses and results from
X-ray and radio data, respectively.  Section~6 is devoted to the
interpretation and discussion of the A523 structure based on
multiwavelength data.

Throughout this paper, we use $H_0=70$ km s$^{-1}$ Mpc$^{-1}$ and
$h_{70}=H_0/(70$ km s$^{-1}$ Mpc$^{-1}$) in a flat cosmology with
$\Omega_0=0.3$ and $\Omega_{\Lambda}=0.7$. In the adopted cosmology,
1\arcm corresponds to $\sim 115$ \kpc at the cluster redshift.  Unless
otherwise stated, we indicate errors at the 68\% confidence level
(hereafter c.l.).

\section{OPTICAL OBSERVATIONS AND DATA SAMPLE}
\label{data}

\subsection{Spectroscopic and photometric observations}
\label{optdata}

Multi-object spectroscopic observations of A523 were carried out at
the TNG in 2012 December and 2014 January. We used the instrument
DOLORES in MOS mode with the LR-B
Grism\footnote{http://www.tng.iac.es/instruments/lrs}. In summary, we
observed six MOS masks for a total of 210 slits. The total exposure time
was 3600 s for three masks, 5400 s for two masks and 7200 s for the
last one.

Reduction of spectra and radial velocities computation with the
cross-correlation technique (Tonry \& Davis \citeyear{ton79}) were
performed using standard {\sevensize IRAF}\footnote{{\sevensize IRAF}
is distributed by the National Optical Astronomy Observatories, which
are operated by the Association of Universities for Research in
Astronomy, Inc., under cooperative agreement with the National Science
Foundation.}  tasks, as done with other clusters included in our DARC
sample (e.g., Boschin et al. \citeyear{bos12}). In two cases (IDs.65
and 110; see Table~\ref{catalogA523}) the redshift was estimated
measuring the wavelength location of emission lines in the
spectra. Our spectroscopic catalogue lists 132 galaxies in the field of
A523.  We corrected the nominal velocity errors provided by the
cross-correlation technique by multiplying them by a factor of 2.2 
as derived from multiple measures of the same targets (e.g., Boschin
et al. \citeyear{bos13}). Taking into account the above correction,
the average value of the $cz$ errors is 72 \kss.

Our photometric observations were carried out with the Wide Field
Camera (WFC), mounted at the prime focus of the 2.5m INT telescope.
We observed A523 in $g$, $r$ and $i$ Sloan-Gunn filters in photometric
conditions and a seeing of $\sim$1.4\arcss.  The WFC consists of a
four--CCD mosaic covering a 33\arcmm$\times$33\arcm field of view
(FOW), with only a 20 per cent marginally vignetted area. We took nine
exposures of 600 s in $g$ filter, and nine frames more of 360 s
exposure in $r$ and $i$ filters, respectively. So a total of 5400 s in
$g$ filter, and 3240 s in $r$ and $i$ bands. Details on observation
procedures and data reduction are described in Barrena et
al. (\citeyear{bar07}).  As a final step, we corrected the Galactic
extinction $A_g=0.569$, $A_r=0.386$ and $A_i=0.276$ following values
listed by NED\footnote{NASA/IPAC Extragalactic Database which is
operated by the Jet Propulsion Laboratory, California Institute of
Technology, under contract with the National Aeronautics and Space
Administration.}.  We estimated that our photometric sample is
complete down to $g=22.4$ (23.4), $r=21.4$ (22.7) and $i=21.2$ (22.4)
for signal-to-noise ratio $S/N=5$ (3) within the observed field.

\subsection{Spectroscopic catalogue  and notable galaxies}
\label{optcat}

\begin{figure*}
\centering 
\includegraphics[width=18cm]{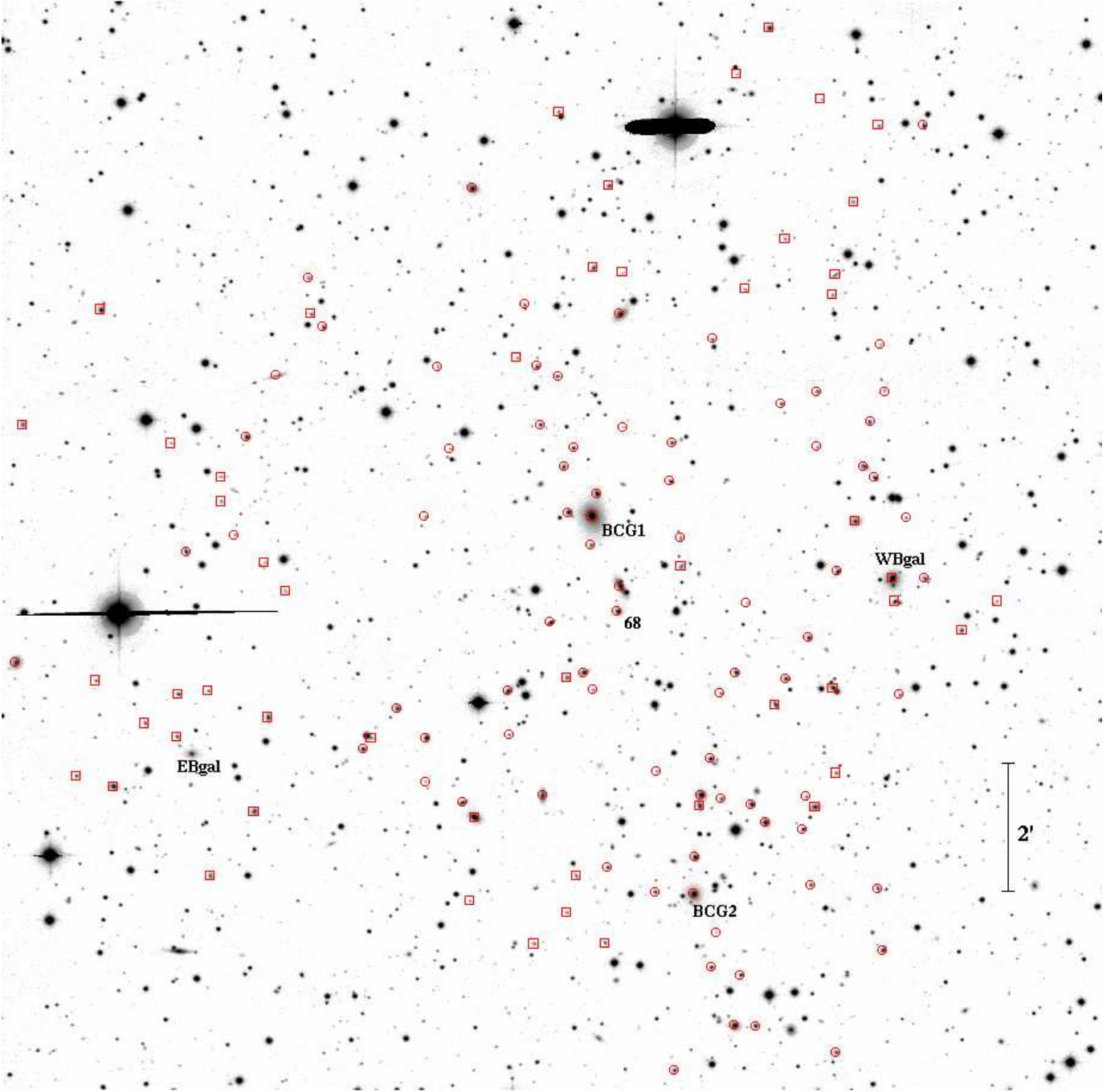}
\caption{INT $r$-band image of Abell 523. Open symbols
highlight galaxies of our spectroscopic catalogue.  Circles and squares
indicate cluster members and non-member galaxies, respectively (see
Table~\ref{catalogA523}). Labels refer to specific galaxies discussed
in the text. North is up and east is left. The FOW of the image is $2\times2$ \hh.}
\label{figottico}
\end{figure*}

Table~\ref{catalogA523} lists the velocity catalogue (see also
Fig.~\ref{figottico}): identification number of each galaxy, ID
(Col.~1); right ascension and declination, $\alpha$ and $\delta$
(J2000, Col.~2); (dereddened) $r$ magnitude (Col.~3);
heliocentric radial velocities, $V=cz_{\sun}$ (Col.~4) with errors,
$\Delta V$ (Col.~5). With the exception of one galaxy, INT dereddened
magnitudes are available.

\input{catalogA523a1.tex}

\begin{figure}
\centering 
\includegraphics[width=7cm]{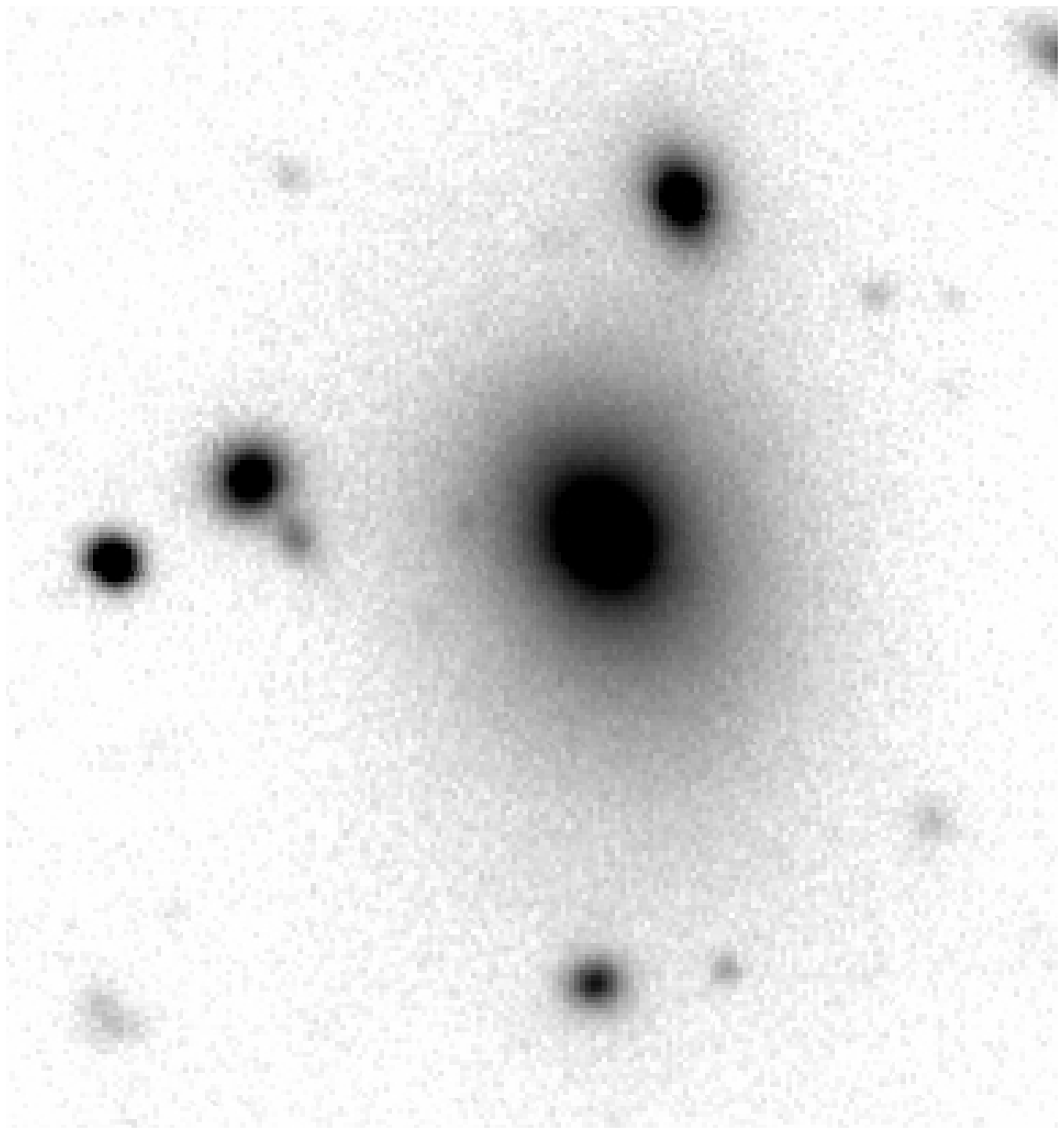}
\caption{Zoomed-in INT  $r$-image to view the details of the 
halo of BCG1, the brightest cluster galaxy specifically related to the
northern subcluster.  (The FOW of the image is $\sim 125\times
130$ \kpcc).}
\label{figBCG1}
\end{figure}

The cluster galaxy population is dominated by the galaxy ID~75 (BCG1)
lying in the northern region. BCG1 also shows an important, diffuse
halo slightly elongated towards SSW (see Fig.~\ref{figBCG1}).  There
is no obvious discrete X-ray/radio source associated with this
galaxy. The second brightest galaxy ID~56 (BCG2) lies in the southern
region. BCG2 is 1.1 mag fainter than the brightest galaxy (BCG1) and
only 0.7-0.8 mag brighter than other luminous cluster galaxies.

G11 detected nine discrete radio sources in the field of
A523 (see their fig.~2), two of them having a measured redshift. The
prominent head--tail radio galaxy in the central region of the cluster
is the member galaxy ID~68, while in the South a discrete radio source is
associated to the cluster member ID~56 (BCG2).  Among the discrete radio
sources listed by van Weeren et al. (\citeyear{vanwee11a}; see their
fig.~6), their source A is the head--tail galaxy cited above, source B
is a probable background galaxy (by visual inspection of our INT
images) and source C is a pointlike source (quasar?) with unknown
redshift.

\section{ANALYSIS OF THE OPTICAL DATA}
\label{opt}

\subsection{Member selection and global properties}
\label{memb}

\begin{figure}
\centering
\resizebox{\hsize}{!}{\includegraphics{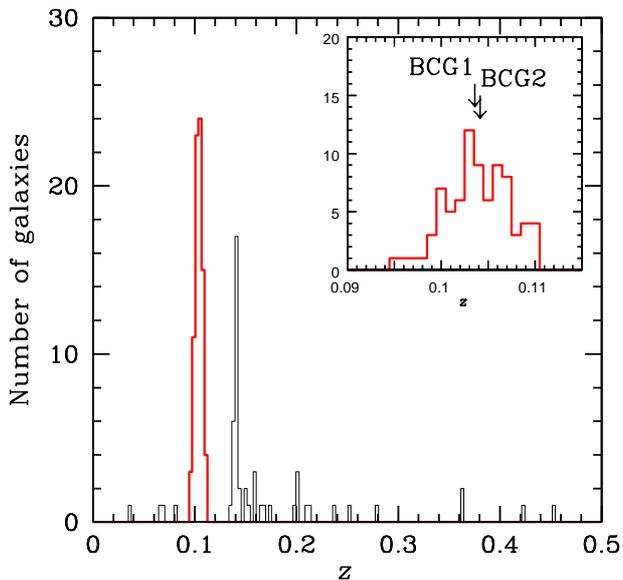}}
\caption
{Redshift galaxy distribution. Thick red line histogram refers to the
  80 galaxies assigned to A523 according to the 1D-DEDICA
  reconstruction method. An important background structure at $z\sim
  0.14$ (BACKstruct) is also evident in the figure. The inset figure
  shows the member-galaxy distribution with the indication
  of the BCG1 and BCG2 velocities.}
\label{fighisto}
\end{figure}

As usual in the analysis of DARC clusters, the selection of cluster
members was performed in two steps. First, we run the 1D version
of the adaptive-kernel method by Pisani (\citeyear{pis93}; hereafter
1D-DEDICA) on the 132 galaxies of our spectroscopic catalogue.  
DEDICA is a non-parametric, scale-independent method for density
reconstruction.  The output of this method gives a list of groups with
related statistical significance and their members.  The estimate of
the (Gaussian) kernel sizes is done through and iterative, optimized
procedure.  In the case of A523, the 1D-DEDICA method detects the
cluster as a peak at $z\sim0.103$ populated by 80 galaxies considered
as fiducial cluster members (in the range $28\,498\leq V
\leq 33\,072$ \kss, see Fig.~\ref{fighisto}). The 52 non-members are 4
foreground and 48 background galaxies, respectively. In
particular, a rich, background peak lies at $z\sim0.140$ with
26 assigned galaxies (hereafter BACKstruct).

\begin{figure}
\centering
\resizebox{\hsize}{!}{\includegraphics{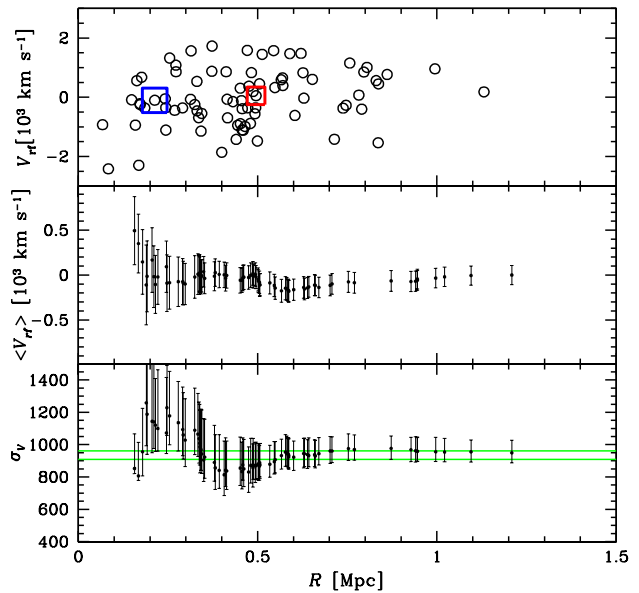}}
\caption
{Top panel: rest-frame velocity versus projected clustercentric
  distance for the 80 member galaxies. The very large blue square
  indicates BCG1 and the large red square indicates BCG2.  
Middle and bottom panels: integral profiles of mean rest-frame
  velocity and LOS velocity dispersion, respectively.  The mean (and
  dispersion) at a given (projected) radius $R$ from the cluster
  center is estimated by considering all galaxies within that radius
  -- the first value computed on the five galaxies closest to the
  center. The $68\%$ error bands are also shown.  Reference values for
  global mean velocity and velocity dispersion are
  $\left<V\right>=31\,165\pm107$ \ks and $\sigma_{\rm
  V}=949_{-60}^{+80}$ \kss.  In the bottom panel, the horizontal,
  green line represents the 1sigma error band of the X-ray temperature
  ($kT_{\rm OUT}=5.3\pm0.3$ keV, see Sect.~\ref{xspec}) and converted
  in $\sigma_{\rm V}$ assuming energy density  equipartition
  between ICM and galaxies, i.e. $\beta_{\rm spec}=1$, see
  Sect.~\ref{disc}.  }
\label{figprof}
\end{figure}

In a second step, we combined the spatial and velocity information by
running the ``shifting gapper'' method based on a gap of 1000 \ks
(Fadda et al. \citeyear{fad96}; Girardi et al. \citeyear{gir96}).
The ``shifting gapper'' method rejects galaxies that are too far
in velocity from the main body of galaxies within a fixed bin that
shifts along the distance from the cluster center. We considered the
standard values of the velocity gap, 1000 \ks in the cluster rest
frame, and of the radial bin, 0.6 \hh.  The determination of the
center is not obvious in the case of an unrelaxed cluster.  In the
case of A523 we considered the biweight position in R.A. and Dec. of
the fiducial member galaxies
[R.A.=$04^{\mathrm{h}}59^{\mathrm{m}}08\dotsec97$, Dec.=$+08\degree
48\arcmm 06\dotarcs3$ (J2000.0)]. The biweight is a robust
statistic for determining the central location of a distribution.  It
was suggested by Tukey (\citeyear{tuk60}) as an improvement for
non-Gaussian and contaminated normal distributions and requires an
auxiliary scale estimator, the median absolute deviation from the
sample median as used in the ROSTAT statistical routines\footnote
{http://bima.astro.umd.edu/nemo/man\_html/rostat.1.html} (Beers et
al. \citeyear{bee90}). We also considered the X--ray center listed
by B\"ohringer et al.  (\citeyear{boe00}), and the density peak of the
\chandra\ X-ray surface brightness as determined in the present study.
Independently of the adopted center, the ``shifting gapper'' procedure
confirms the 80 fiducial cluster members in the velocity peak.
Hereafter we adopt the biweight center.  The projected phase space of
member galaxies is shown in Fig.~\ref{figprof} (top panel).

The analysis of the velocity distribution of the 80 cluster members
was performed by using the biweight estimators of location and scale
as implemented in the ROSTAT statistical routines (Beers et
al. \citeyear{bee90}).  Note that Beers et al. (\citeyear{bee90})
tested the resistance, robustness, and efficiency of these and other
estimators with extensive simulations for a number of common cases
realizable in samples of cluster velocities.  Following that study and
Girardi et al. (\citeyear{gir93}), we always used the biweight
estimator of scale for samples of $n\ge 15 $ galaxies and the gapper
estimator for samples of $n< 15 $ galaxies.  In the case of A523, we
calculated the mean cluster redshift, $\left<z\right>=0.1040\pm0.0004$
(i.e., $\left<V\right>=31\,165\pm107$ \kss), and the global
line-of-sight (LOS) velocity dispersion, $\sigma_{\rm
V}=949_{-60}^{+80}$ \kss. The robustness of these estimates with
respect to the cluster radius is confirmed by our analysis of the mean
velocity and velocity dispersion profiles (see Fig.~\ref{figprof}).

\subsection{Cluster substructure}
\label{sub}

We analysed the presence of substructure on the basis of the velocity
distribution of galaxies, their projected positions on the sky,
and combining these two pieces of information (1D, 2D, and 3D tests).
According to the analysis of the velocity distribution, there is no evidence for
possible deviations from Gaussianity according to a variety of
parameters (kurtosis, skewness, tail index, and asymmetry index; Bird
\& Beers \citeyear{bir93}).

\begin{figure}
%\centering
\resizebox{\hsize}{!}{\includegraphics{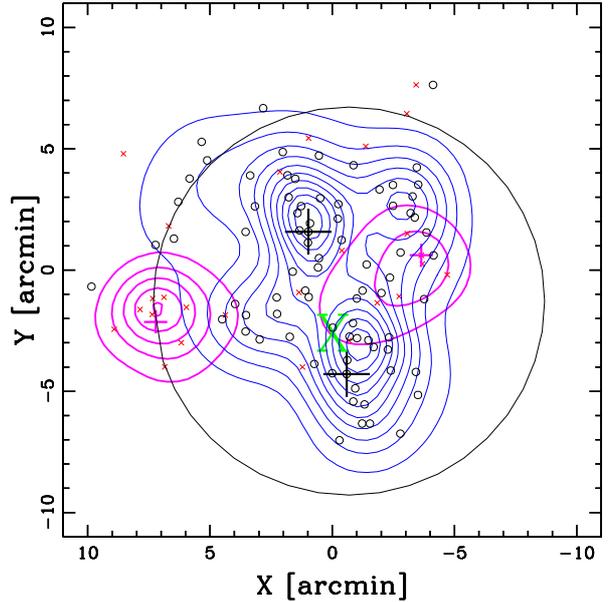}}
\caption{Spatial distribution on the sky of
  the 80 spectroscopic cluster members and relative isodensity contour
  map obtained with the 2D-DEDICA method (small black open circles and
  thin blue countours). The two huge (black) crosses indicate the
  position of BCG1 and BCG2 which are close to the NNE and SSW galaxy
  density peaks, respectively.  A third, less dense peak lies in the
  NW.  Thick magenta contours and small red crosses refer to the
  BACKstruct galaxies and large magenta crosses indicate EBgal and WBgal,
  suggested to be two dominant galaxies in the ESE-Bgroup and
  WNW-Bgroup.  The plot is centred on the cluster center and the
  circle contains the cluster within a radius equal to 8\arcm$\sim
  0.92$ \h ($\sim$ $0.5R_{200}$).  The large (green) 'X' label indicates the
  peak of the X-ray surface brightness as determined in this study.}
\label{figk2z}
\end{figure}

We analysed the spatial distribution of the 80 spectroscopic member
galaxies by using the 2D adaptive-kernel method of Pisani et
al. (\citeyear{pis96}, hereafter 2D-DEDICA, see also the Appendix in
Girardi et al. \citeyear{gir96}).  Our results are presented in
Fig.~\ref{figk2z} and Table~\ref{tabdedica2dz} and highlight a
SSW-NNE elongated structure with two very significant galaxy
peaks. BCG1 and BCG2 are the two dominant galaxies of the northern and
southern subclusters, respectively.  We also detected a third, low
density peak, at the limit of our detection threshold, in the NW.
However, our spectroscopic data not cover the entire cluster
field and are affected by magnitude incompleteness due to unavoidable
constraints in the design of the MOS masks.  The analysis of the
photometric catalogues can offer an unbiased description of the 2D
galaxy distribution. On the other hand, in the specific case of A523,
the presence of the BACKstruct at $z\sim 0.14$ has to be taken into
account (Sect.~\ref{back}) and the analysis of the INT photometric
catalogues is given in Sect.~\ref{sub2D}.

\input{tabdedica2dz.tex}

\begin{figure}
\centering 
\includegraphics[width=8cm]{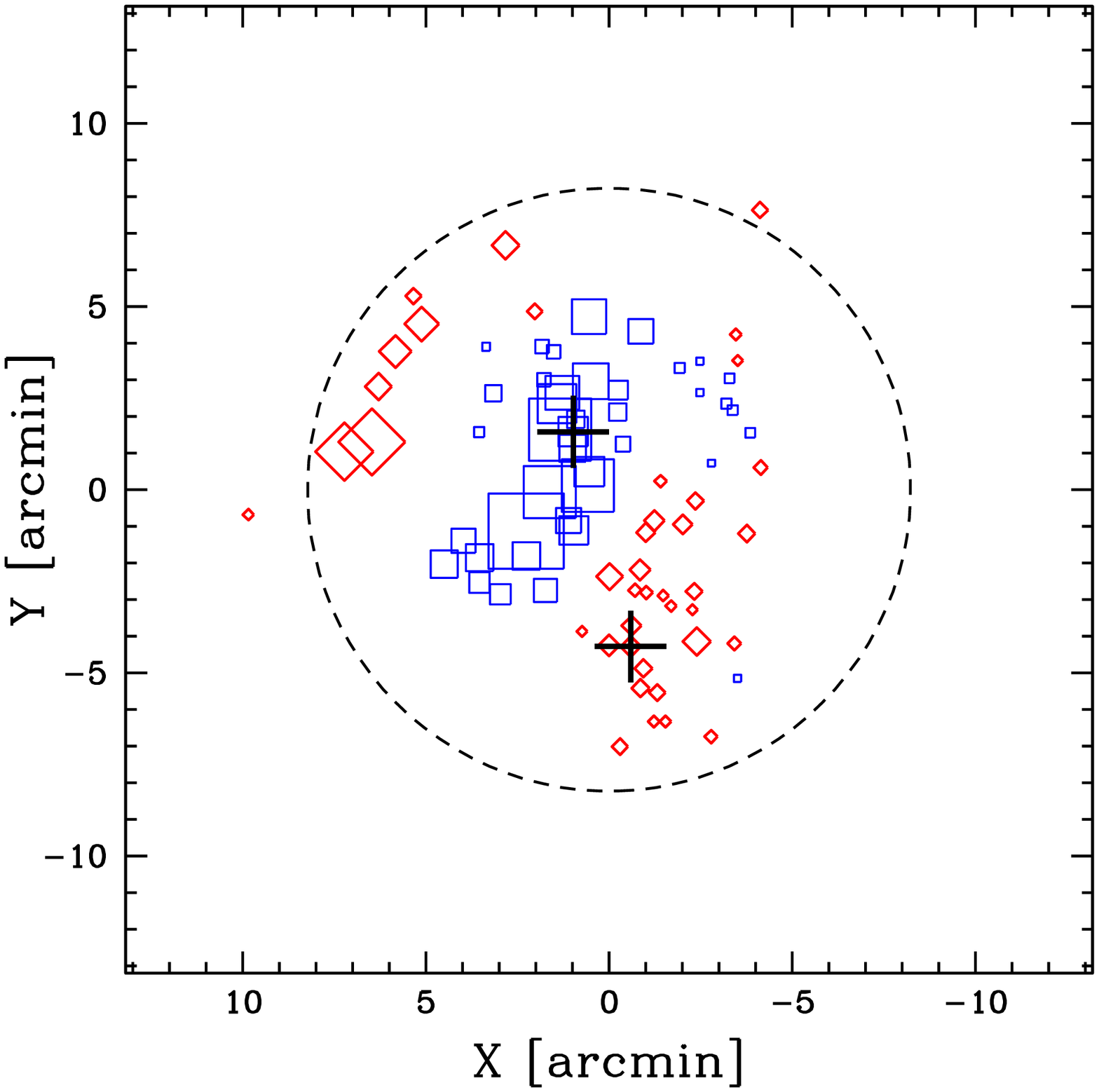}
\includegraphics[width=8cm]{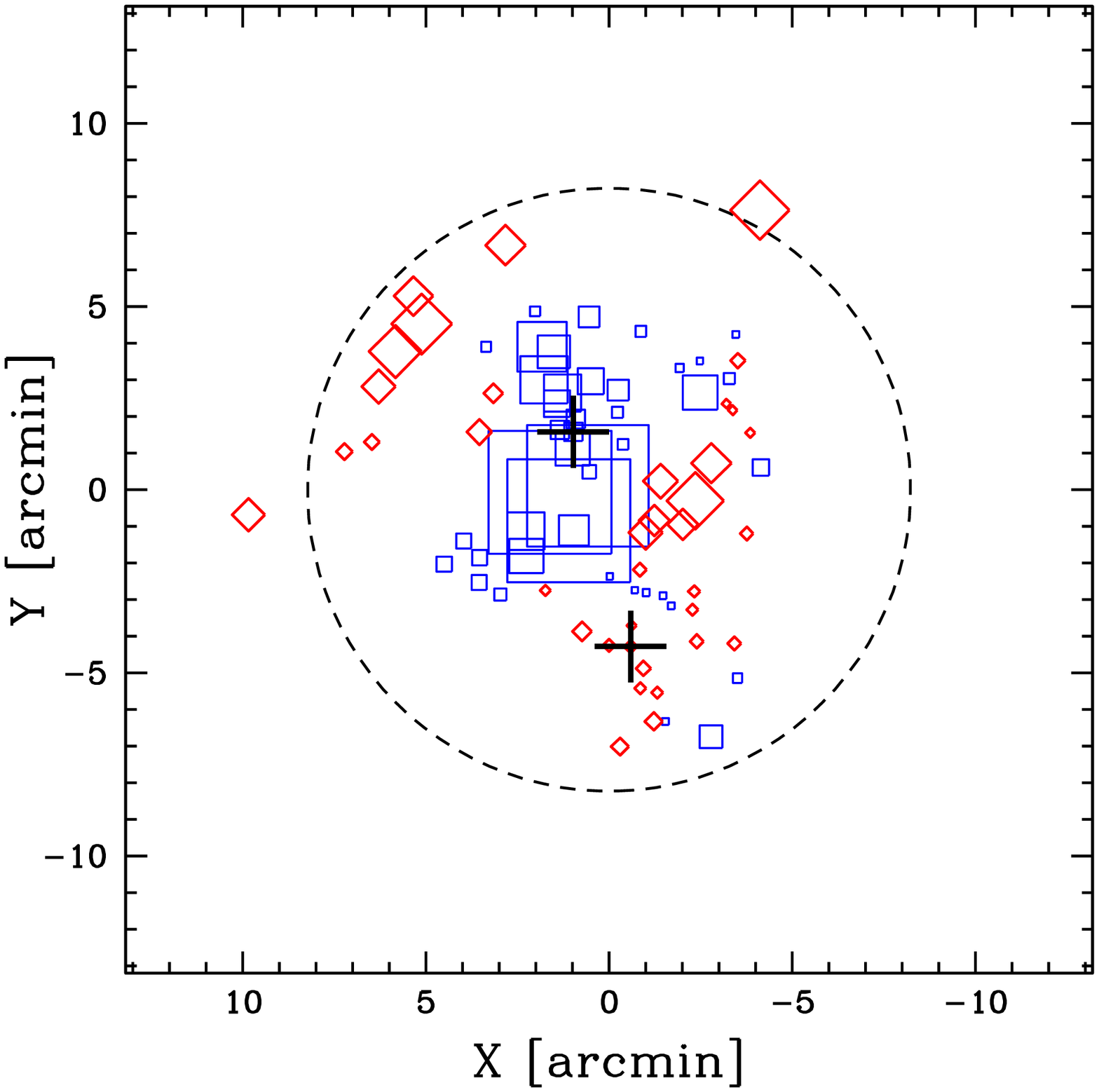}
%\resizebox{\hsize}{!}{\includegraphics{figdssegno10v.eps}}
%\resizebox{\hsize}{!}{\includegraphics{figdssegno5v.eps}}
\caption
{Dressler \& Schectman bubble-plot for the $\Delta$-test based on
  local mean velocities estimated on $N_{\rm nn}+1$ galaxies. Spatial
  distribution of the 80 cluster members, each marked by a symbol: the
  larger the symbol, the larger is the deviation of the local mean
  velocity from the global mean velocity with scale $\propto$
  exp(|$\delta_{i,V}$|).  Thin blue squares and thick red rotated
  squares show where the local value is smaller or larger than the
  global value.  The plot is centred on the cluster center and the
  circle (dashed line) contains the region within a radius equal to
  8\arcm $\sim 0.92$ \hh. Upper and lower panels refer to the cases
  $N_{\rm nn}=10$ and $N_{\rm nn}=5$, respectively.}
\label{figds10v}
\end{figure}

As for the 3D analysis, we searched for a correlation between velocity
and position information checking for a velocity gradient (see, e.g.,
den Hartog \& Katgert \citeyear{den96} and Girardi et
al. \citeyear{gir96}) and applied the $\Delta$-test by Dressler \&
Schectman (\citeyear{dre88}).  The $\Delta$-test is a powerful
test for 3D substructure (Pinkney et al.  \citeyear{pin96}).  We also
used two kinematical estimators alternative to the $\delta_{i}$
parameter of the standard $\Delta$-test (Barrena al. \citeyear{bar11},
see also Girardi et al. \citeyear{gir97}). We considered separately
the contribution of the deviation of the local mean velocity from
global mean velocity $\delta_{i,V}= [(N_{\rm
nn}+1)^{1/2}/\sigma_V](V_{\rm loc} -
\left<V\right>)$ and the contribution of the deviation of the local
velocity dispersion from the global velocity dispersion $\delta_{i,
  {\rm s}}= [(N_{\rm nn}+1)^{1/2}/\sigma_V](\sigma_{V,{\rm loc}} -
\sigma_V)$, where the subscript loc denotes the local quantities
computed in the group containing the $i$-th galaxy and its
$N_{\rm{nn}}=10$ neighbours. In all the above tests, the statistical
significance is based on 1000 Monte Carlo simulated clusters obtained
shuffling galaxies velocities with respect to their positions. In
A523 we found no significant evidence for a velocity gradient and
substructure.  However, in Fig.~\ref{figds10v} we show the Dressler \&
Schectman bubble-plots resulting from the indicator $\delta_{i,V}$ for
$N_{\rm nn}=10$ and $N_{\rm nn}=5$.  In fact, these plots are quite
suggestive of how BCG1 is surrounded by galaxies having lower velocity
than galaxies surrounding BCG2, in agreement with our following
analyses.

\begin{figure}
\centering 
\resizebox{\hsize}{!}{\includegraphics{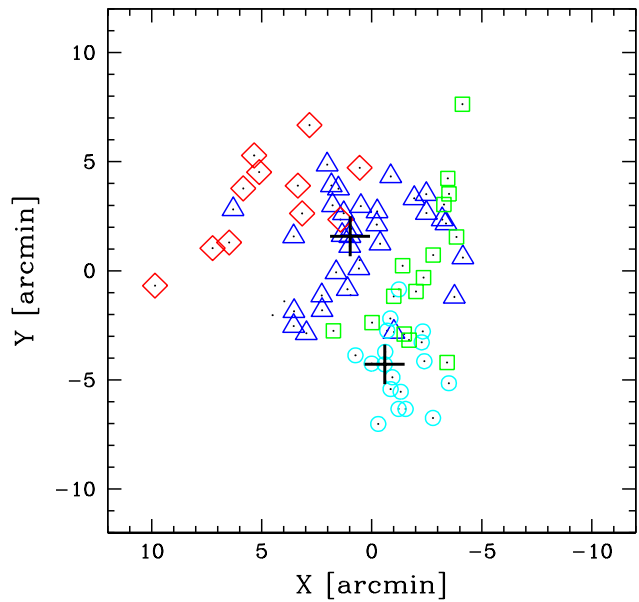}}
\caption
{Spatial distribution of the 80 cluster members (small black points)
within a region of $\sim 2.8 \times 2.8$ \h centred on the
cluster center, where the galaxies of the four subclusters detected
in the 3D-DEDICA analysis are indicated by large symbols. Blue
triangles, cyan circles, green squares, and red rotated squares
indicate galaxies of groups with increasing mean velocities. The group
formed of galaxies indicated by cyan circles is the closest one to
the mean cluster velocity.  The members of the two main subclusters
are indicated by blue triangles (the northern one) and cyan circles
(the southern one).  The northern subcluster stands out for its low
velocity being surrounded by two high velocity groups.  The two
crosses indicate the two BCGs.  }
\label{figdedica3D}
\end{figure}

We performed the 3D-DEDICA analysis detecting two very significant
density peaks corresponding to the northern and southern subclusters
and having peak velocities of 30\,932 and 31\,121 \kss. We also
detected two other peaks, still $>99\%$ significant, characterized by
very high velocities (32\,024 and 32\,806 \kss). Due to the problem of
combining  non-homogeneous quantities like position and velocity, one
should be cautious in treating with results coming from methods of 3D
reconstruction.  However, in the case of A523, both the 3D-DEDICA
analysis and the Dressler \& Schectman bubble-plot agree in showing a
structure more complex than the bimodal one and in finding that the
low velocity galaxy population around BCG1 is surrounded by two
populations at high velocity (cfr. Fig.~\ref{figdedica3D} and the
lower panel in Fig.~\ref{figds10v}).  This complexity is likely the
reason why we detected no significant evidence for a velocity gradient
and for substructure using classic tests and suggests the need of a
larger redshift sample to successfully describe the cluster velocity
field.

\begin{figure}
\centering
\resizebox{\hsize}{!}{\includegraphics{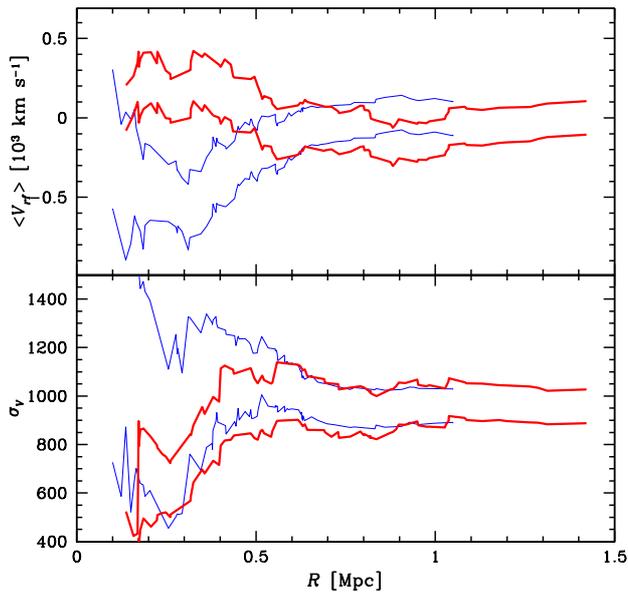}}
\caption
{Error bands of mean velocity and velocity dispersion profiles
  computed as in middle and bottom panels of Fig.~\ref{figprof}. Thin
  blue lines refer to BCG1 as center and thick red lines refer
  to BCG2 as center.  The comparison highlights the difference in mean
  velocity between populations surrounding the two BCGs.}
\label{figprofNS}
\end{figure}

Although the northern and southern subclusters cannot be separately
detected in the velocity distribution, we obtained three alternative
estimates of their relative LOS velocity, all indicating that the
northern subcluster, related to BCG1, is populated by galaxies having
on average a velocity lower than galaxies populating the southern subcluster,
related to BCG2. We considered (i) the LOS velocity difference of
the BCGs in the cluster rest frame, $\Delta V_{\rm rf}\sim 150$ \kss,
and (ii) the difference between the two main subsystems detected in
the 3D-DEDICA analysis, $\Delta V_{\rm rf}\sim 170$ \kss.  The (iii)
estimate is based on the analysis of the mean velocity profiles
computed using BCG1 or alternatively BCG2 as centres.
Fig.~\ref{figprofNS} (upper panel) shows that as the galaxy
populations surrounding BCG1 and BCG2 within 0.3 \h do differ in their
mean velocity, while when considering galaxies at larger radii the
mixing of the populations makes the values of the mean velocity to
converge to the global value. Considering galaxies at $R\siml 0.3$
\hh, we computed $\left<V_{\rm N}\right>=30\,561\pm168$ \ks and
$\left<V_{\rm S}\right>=31\,283\pm137$ \kss, which means a velocity
difference significant at the 2.7 sigma c.l. ($\Delta V_{\rm rf}\sim
650$ \kss).

We also performed the analysis of the velocity dispersion profiles
(Fig.~\ref{figprofNS}, lower panel).  As for the population around
BCG2, galaxies at $R\siml 0.3$ are characterized by a low value of
velocity dispersion, while at larger radii the profile increases
likely due to the contamination of the northern subcluster.  Thus we
can estimate $\sigma_{\rm V,S}\sim 650$ \ks for the galaxy population
related to BCG2.  The population around BCG1 is characterized by a
higher nominal value of velocity dispersion, but the huge uncertainty
prevents us to give an estimate of $\sigma_{\rm V,N}$. 

\subsection{The background galaxy structure at \texorpdfstring{$z\sim 0.14$}{}}
\label{back}

The BACKstruct is not a dense galaxy system, but rather a quite sparse
structure.  Applying the 2D-DEDICA analysis to the BACKstruct galaxies
we detected two loose overdensities tracing the ESE-WNW direction
(ESE-Bgroup and WNW-Bgroup, see Table~\ref{tabdedica2dback}).
Fig.~\ref{figk2z} shows the relevant contour map (red contours).

The galaxy population of the WNW-Bgroup is dominated by the galaxy
ID~8 (WBgal), while the galaxy population of the ESE-Bgroup seems to
be related with the bright galaxy located at
R.A.=$04^{\mathrm{h}}59^{\mathrm{m}}38\dotsec2$, Dec.=$+08\degree
45\arcmm 58\arcs$ (J2000.0; EBgal).  X-ray emission is associated with
both WBgal and EBgal and radio emission with EBgal only (see
Fig.~\ref{figimage}).  As for the WBgal galaxy listed in
Table~\ref{catalogA523}, its magnitude refers to a close couple of
galaxies/nuclei in the same light halo, and the spectrum refers to the
northern galaxy/nucleus. As for the EBgal, this is not listed in our
redshift catalogue. However, its colour is typical of that of the
background structure and is surrounded by a faint halo and a few small
galaxies.

\input{tabdedica2dback}

In the effort of understanding the nature of the BACKstruct, we note that
A523 is part of the supercluster SCL62 (Einasto et
al. \citeyear{ein01}) together with Abell 515, 525, 529, 532 in the
range of photometric redshifts $z_{\rm phot}$=0.04-0.14 (Gal et
al. \citeyear{gal00}).  The BACKstruct might be related to the
outskirts, at $\sim 6$ \hh, of Abell 525 which is the closest to A523
in the plane of the sky and has $z_{\rm phot}\sim 0.14$.

\subsection{2D cluster structure based on the photometric data}
\label{sub2D}

\begin{figure}
\centering
\resizebox{\hsize}{!}{\includegraphics{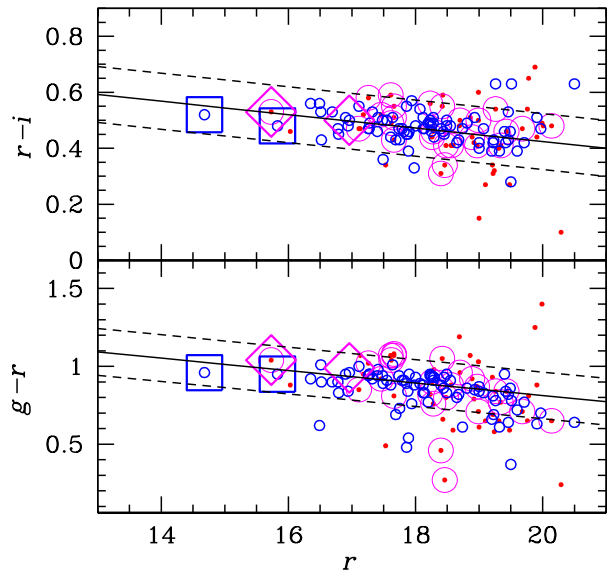}}
\caption
{Upper and lower panels: $r$--$i$ versus $r$ and $g$--$r$ versus $r$
    diagrams, respectively. Galaxies with available spectroscopy are
    shown by blue open circles (cluster members) and small red dots
    (field members).  Very large, blue open squares indicate BCG1 and BCG2.
    In both panels the solid line gives the best--fit CMR as
    determined on spectroscopic cluster members; the dashed lines
    define the regions within which photometric cluster members are
    selected.  To stress the problem of contamination in A523 due to
    the background field, we indicate galaxies belonging to the
    BACKstruct with large magenta open circles.  Very large, magenta
    open rotated squares indicate WBgal and EBgal, the
    second one not included in our spectroscopic catalogue.}
\label{figcm}
\end{figure}

Here we present our results about the cluster structure based on the
INT photometric catalogues.  We selected photometric members on the
basis of both ($r$--$i$ versus $r$) and ($g$--$r$ versus $r$)
colour-magnitude relations (CMRs), which indicate the early-type
galaxies locus.  The equations of the two CMRs are
$r$--$i$=0.904-0.024$\times r$ and $g$--$r$=1.613-0.004$\times r$ and
we selected galaxies within a colour range of 0.1 and 0.15,
respectively (see Fig.~\ref{figcm}). To determine the CMRs we
applied a 2$\sigma$-clipping fitting procedure to the spectroscopic
members (see Boschin et al. \citeyear{bos12} and refs. therein).  In
particular, to reduce the contamination of non-member galaxies, we
considered the galaxies with $r\le 19$, corresponding to $\sim 2$ mag
after $M^*$, for a total of 196 members in the whole photometric
catalogue.

\begin{figure}
\centering
%\resizebox{\hsize}{!}{\includegraphics{figk2cm19nostarradio.eps}}
\includegraphics[width=12cm]{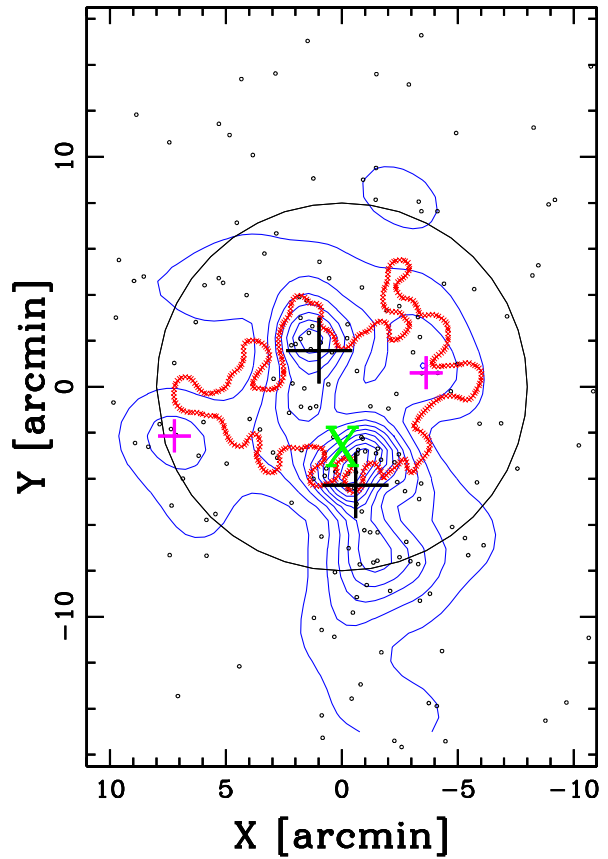}
\caption{Spatial distribution on the sky of the photometric cluster
  members with $r\le 19$ (small black dots). The contour map is
  obtained with the 2D-DEDICA method (blue contours).  The two large
  black crosses indicate the position of BCG1 and BCG2 and the two
  magenta crosses indicate EBgal and WBgal, likely the dominant
  galaxies in the background structure (see Fig.~\ref{figk2z} and the
  text).  The radio emission is shown in a schematic way (small red
  crosses) to give the lowest contour of Fig.~\ref{radioimage} (middle
  panel).  The large green 'X' label indicates the peak of the X-ray surface
  brightness.  The plot is centred on the cluster center and the
  circle contains the region within a radius equal to 8\arcm 
  $\sim 0.92$ \hh.}
\label{figk2}
\end{figure}

The results of the 2D-DEDICA analysis are presented in
Fig.~\ref{figk2} and Table~\ref{tabdedica2d} and confirm the results
obtained in the spectroscopic sample, that is the presence of the
northern and southern subclusters, NNE(2D) and SSW(2D) peaks. In
addition, we detected S-SSW(2D), a less dense peak in the southern
region, poorly sampled by our spectroscopic catalogue, tracing the
SSW-NNE direction too. We also detected two minor density peaks, ESE(2D)
and WNW(2D), tracing the ESE-WNW direction, but they might be
partially or totally spurious, as due to the BACKstruct contamination,
as discussed in the following.

\input{tabdedica2d.tex}

Fig.~\ref{figcm} shows that several galaxies of the BACKstruct have
colours similar to those of A523 red sequence galaxies thus likely
contaminating our catalogue of photometric members. In fact, 21 out of
the 26 galaxies of the BACKstruct are (erroneously) classified as
photometric members of A523. Therefore our photometric sample is
likely contaminated by the BACKstruct galaxies, in particular in the
WNW and ESE regions.  The ESE(2D) peak, located in a region where we
do not detect any cluster overdensity, might be not related to A523,
while the WNW(2D) peak, located in a region where both the WNW-Bgroup
of the BACKstruct and the NW(z) of the A523 cluster are projected,
might be contaminated (cfr. Fig.~\ref{figk2z} and Fig.~\ref{figk2}).
As for the SSW-NNE region traced by the two main subclusters, we
estimated a small ($\siml 15\%$) contamination. This estimate is the
ratio of the number of the BACKstruct galaxies classified as A523
photometric members to the number of A523 galaxies (spectroscopic
sample) classified as A523 photometric members. Therefore, we expect a
small effect in the estimate of the 2D positions of the peaks of the
northern and southern subclusters as listed in Table~\ref{tabdedica2d}
(see also the good agreement with those in Table~\ref{tabdedica2dz}).

\section{X-RAY MORPHOLOGICAL AND SPECTRAL ANALYSIS}
\label{xray}

\subsection {Observation and data reduction}
\label{xraydata}

A523 was observed by \chandra\ with the \acisi\ configuration in
VFAINT mode on 2013, November 04 (obsID 15321). The data were analysed
with \ciao\ 4.6 in conjunction with the \chandra\ Calibration Database
(CALDB) 4.6.1.  All data were reprocessed from the level=1 event files
following the standard \chandra\ reduction threads. In particular we
ran the task {\ttfamily acis\_process\_events} to flag background
events that are most likely associated with cosmic rays and removed
them.  With this procedure, the ACIS particle background can be
significantly reduced compared to the standard grade selection. The
data was filtered to include only the standard event grades 0, 2, 3, 4
and 6.  From low surface brightness regions of the active chips we
extracted light curves which were cleaned from soft proton flares
using the \ciao\ task {\ttfamily deflare} with the {\ttfamily clean}
option. The observation was quiescent resulting in a negligible loss
of time and in an effective exposure time of 27 ks. Point source
detection was performed using {\ttfamily wavedetect} on a 0.5-7 keV
band image, supplied with a 1.4 keV psf map to minimize spurious
detections at chip boundaries.  The detection threshold was set to
$10^{-6}$ and the scales were set to a factor of 2 from 1 to 8.  Point
sources were removed using appropriate elliptical regions containing
99\% of their flux.  Spectral extraction was performed with the
\ciao\ tool {\ttfamily specextract} creating the appropriate response
files, RMF and ARF, for the selected spatial region.  Spectral fitting
was performed with \xspec\ 12.8.2 using the C-statistic.  We rebinned
the spectra to ensure at least 20 photons per bin to aid in model
selection and computational speed.  The energy range was restricted to
0.5-7.0 keV and all spectra were fitted including components for the
cluster emission and background. We used a single thermal APEC
component to model the ICM emission modified by Galactic absorption
fixed at $1.2\times10^{21}$ cm$^{-2}$ (Dickey \& Lockman 1990, a
similar value $1.05\times10^{21}$ cm$^{-2}$ is reported by Kalberla et
al.  \citeyear{kal05}).  To account for the background we included
additional spectral components in the fits: we included two APEC
components ($kT$ = 0.07 keV and 0.2 keV, the former un-absorbed) to
account for the Galactic foreground and a power law component
($\Gamma$ = 1.41) for the Cosmic X-ray Background due to unresolved
AGNs. To account for the instrumental background, we followed the
analytical model of Bartalucci et al. (\citeyear{barta14}) which was
not folded through the ARF. The background parameters were constrained
by fitting spectra extracted from regions free of cluster emission at
the edges of the FOW, and then the fitted normalizations
were rescaled accordingly to the source extraction area.

\subsection {X-ray image and surface brightness analysis}
\label{ximage}

\begin{figure}
\centering
%\resizebox{\hsize}{!}{\includegraphics{imax_paper.ps}}
\resizebox{\hsize}{!}{\includegraphics{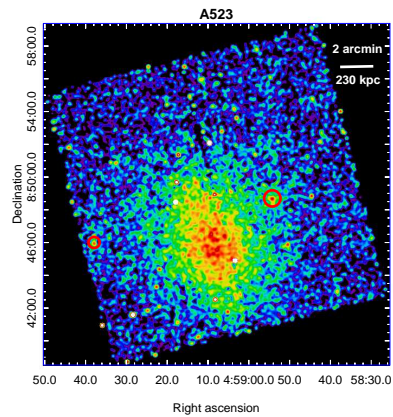}}
\caption
{\chandra\ image in the 0.5--7 keV energy band smoothed on a 10\arcs
  scale. The image was divided by the exposure map at 1.4 keV to
  correct for exposure variations. The detected point sources
  connected with galaxies belonging to the background groups are
  highlighted by two red open circles. 
}
\label{fig:imax}
\end{figure}

We created an image in the 0.5-7 keV band and a corresponding exposure
map at 1.4 keV, to make an exposure corrected image, smoothed on a
scale of 10\arcss, as shown in Fig.\ref{fig:imax}.  We processed the
image to remove the point sources using the CIAO tool {\em dmfilth},
which replaces photons in the vicinity of each point source with a
locally estimated background. We performed a morphological analysis
adopting quantitative measures by applying the power ratios technique
(Buote \& Tsai \citeyear{buo96}), the centroid shifts (Mohr et
al. \citeyear{moh93}) and the concentration parameter (Santos et
al. \citeyear{san08}).  The power ratio method exploits the idea that
the X-ray surface brightness represents the projected mass
distribution of the cluster. The power ratio is a multipole
decomposition of the two dimensional surface brightness inside a given
aperture. We calculated the power ratios within an aperture of
500 \kpc for comparison with previous work. In particular, we used the
$P_3/P_0$ ratio which provides a clear substructure measure
following Cassano et al. (\citeyear{cas10}).  Centroid shifts indicate
that the center of mass of the X-ray emitting gas varies with radius.
Centroid shifts and power ratios are both capable of identifying
highly disturbed systems or systems with significant, well defined
substructures (Poole et al. \citeyear{poo06}).  Following the method
of Poole et al. (\citeyear{poo06}), the centroid shift was computed in
a series of circular apertures centred on the cluster X-ray peak. The
radius of the apertures was decreased in steps of 5\% from 500 \kpc to
25 \kpcc.  The concentration parameter, $c$, is defined as the ratio
of the peak (calculated within 100 \kpcc) over the ambient surface
brightness (calculated within 500 \kpcc). The concentration parameter
differentiates clusters with a compact core (core not disrupted by a
recent merger event) from clusters with a spread distribution of gas
in the core (core disrupted by a recent merger episode).

We applied these techniques and the results are reported
in Table~\ref{tabxmorpho}. All the morphological indicators
confirm quantitatively the simple visual indication of a disturbed
cluster affected by a merger.

\input{tabxmorpho}

\begin{figure}
\centering
\resizebox{\hsize}{!}{\includegraphics{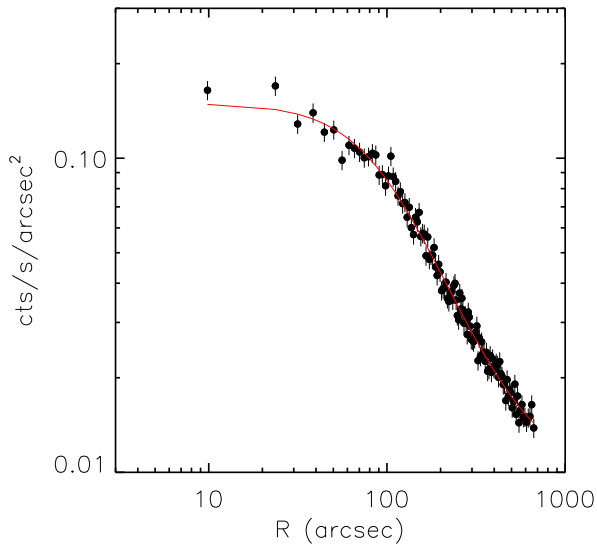}}
\caption
{Surface brightness profile of the X-ray emission of A523. The best
  fit $\beta$-model is also shown in red.}
\label{fig:sbprofile}
\end{figure}

We extracted the surface brightness profile from the exposure corrected
image in the 0.5-7 keV band and we accounted for the X-ray background
by including a constant background component.  The data were grouped to
have 200 counts per bin and the $\chi^{2}$ statistics was applied.  The
best-fitting $\beta$-model has a core radius of $r_c = (181\pm17)$ \kpc
(i.e., $98$\arcs $\pm10$\arcss) and $\beta=0.42\pm0.02$ for a
$\chi^{2}$/d.o.f. = 116/115 (see Fig.\ref{fig:sbprofile}).  The peak
of the surface brightness lies at
R.A.=$04^{\mathrm{h}}59^{\mathrm{m}}08\dotsec9$, Dec.=$+08\degree
45\arcmm 31\arcs$ (J2000.0) and the central gas density, as obtained
in a region of 165 \kpcc, is $1.1\times 10^{-3}\rm{cm}^{-3}$ with an
uncertainty of $4\%$.

\subsection {Spectral analysis}
\label{xspec}

We calculated the mean temperature of the cluster as the temperature
obtained with a spectral fit in the region $0.05 R_{180} < r < 0.2
R_{180}$, where $R_{180} = 1780 (kT/5{\rm{keV}})^{1/2} h(z)^{-1}$
kpc with $ h(z)= (\Omega_{\rm{M}} (1+z)^3 + \Omega_{\Lambda})^{1/2}$
using an iterative procedure to calculate $kT$ and $R_{180}$ (Rossetti
\& Molendi \citeyear{ros10}).  We found $kT= 5.3 \pm 0.3$ keV.  The value
of $kT$ thus estimated within the $0.05 R_{180} < r < 0.2 R_{180}$
region, hereafter $kT_{\mathrm OUT}$, is a good proxy for the global
temperature (see fig.~4 of Leccardi \& Molendi \citeyear{lec08} where
$kT_{\rm M}$ is the global temperature).

\begin{figure}
\centering
\includegraphics[width=9cm]{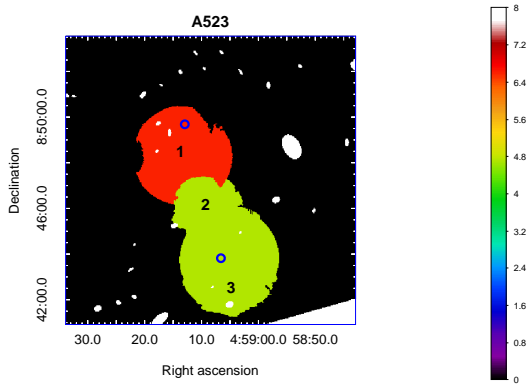}
\caption
{Temperature map (keV) showing that the northern cluster region is
characterized by a higher temperature than the central and southern
regions.  The central 1.5\arcm radius region, indicated by the label
``2'', is centred close to the X-ray peak, precisely at
R.A.=$04^{\mathrm{h}}59^{\mathrm{m}}09\dotsec0$, Dec.=$+08\degree
45\arcmm 55\arcs$ (J2000.0).  The FOW of the image is $\sim
1.5\times 1.5$ \hh.  The two small blue circles indicate BCG1 and
BCG2 in the northern and southern regions (regions 1 and 3,
respectively).}
\label{tmap}
\end{figure}

We  also prepared a coarse two dimensional map of X-ray temperature, using
the contour binning technique of Sanders (\citeyear{san06}) with a
S/N of 50, which resulted in three interesting regions
shown in Fig.~\ref{tmap}. The values for the temperatures are $kT_{\rm
Reg2}=4.7\pm0.4$ keV in the central region, $kT_{\rm Reg1}=6.6\pm0.9$ keV in the northern region, and $kT_{\rm Reg3}=4.7\pm0.5$ keV in the southern region.

\subsection{X-ray luminosity estimate and the \texorpdfstring{$L_{\rm X}$-$T$}{} relation}
\label{lt}

The previously available X-ray information for this cluster is limited
to a \rosat\ All Sky Survey observation which detected the cluster at
a flux of $F_{\rm X}=4.5 \times 10^{-12}$ \fxunits\ in the 0.1-2.4 keV
band, bright enough to be included in the \rosat\ BCS (Ebeling et
al.  \citeyear{ebe98}), with a predicted temperature of 4.3 keV (see
their table~3). Based on the \rosat\ data B\"ohringer et al.
(\citeyear{boe00}, NORAS Survey) list a luminosity
$L_{{\rm X}}(<R_{\rm ap}=0.92 \hhh) = 0.88 \times 10^{44}$
\lxunits\ in the 0.1-2.4 keV band which,  as standardized in the MCXC catalogue
(Piffaretti et al. \citeyear{piffa11}), gives a luminosity of $L_{{\rm
X,500}}= 0.91 \times 10^{44}$ \lxunits\ in the 0.1-2.4 keV band,
whithin the radius\footnote{The radius $R_{\delta}$ is the radius of a
sphere with mass overdensity $\delta$ times the critical density at
the redshift of the galaxy system.} $R_{500}=0.83$ \hh.

This value corresponds to a predicted temperature of
2.9 keV using the best fit $L_{\rm X}$-$T$ relation found in REXCESS
(appendix B of Pratt et al. \citeyear{pratt09}). These predicted
temperature estimates are significantly lower than our measure of the
X-ray temperature based on \chandra\ data suggesting that A523 may be
a significant outlier in the $L_{\rm X}$-$T$ relation (see
Fig.\ref{fig:lt}). To investigate this point we made  our own estimate
of the A523 X-ray luminosity based on \chandra\ data.

\begin{figure}
\centering
\resizebox{\hsize}{!}{\includegraphics[angle=-90]{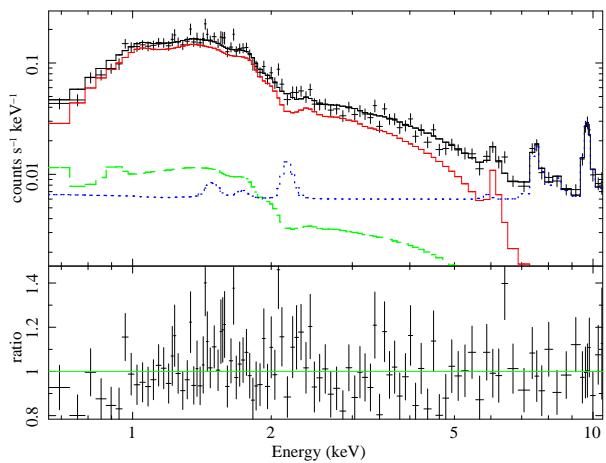}}
\caption
{Upper panel: X-ray spectrum of the source taken from a 337 \kpc
  aperture. The ACIS-I data are shown in black together with the best
  fit model with the cluster component in red (solid line), the
  particle background in blu (dotted line) and the sky background in
  green (dashed line). Lower panel: the ratio of the data over the
  model is also shown.}
\label{fig:spectrum}
\end{figure}

\begin{figure}
\centering
\resizebox{\hsize}{!}{\includegraphics{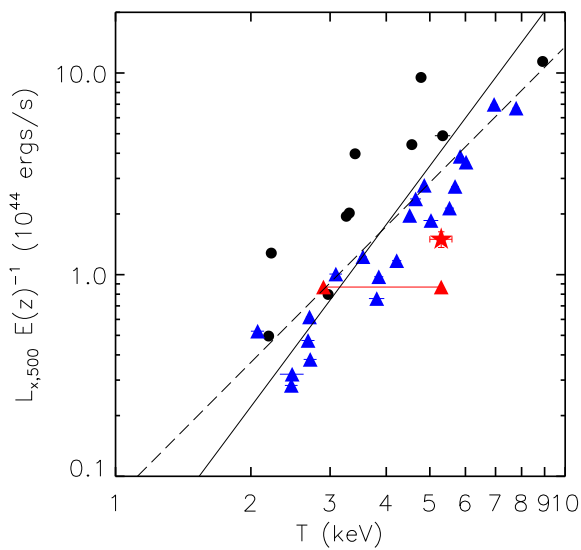}}
\caption
{$L_{\rm X}$-$T$ relation for the REXCESS sample with quantities
  derived from all the emission interior to $R_{500}$ in the 0.1-2.4
  keV band.  Black points represent cool-core systems whereas blue
  triangles represent non cool-core systems according to the
  classification of Pratt et al. (\citeyear{pratt09}).  The best
  fitting power law relation derived from BCES (Y$|$X) (dashed line)
  and BCES orthogonal (solid line) are overplotted.  The two connected
  (red) triangles represent the position of A523 based on the \rosat\
  luminosity estimate and the predicted value of $T=2.9$ keV (triangle
  on the left) and the \chandra\ temperature estimate (triangle on the
  right). The red star represents the position of A523 based on
  the \chandra\ temperature and luminosity estimates.  }
\label{fig:lt}
\end{figure}

For the luminosity calculation we extracted a spectrum from the entire
aperture of radius $0.2 R_{180}$ corresponding to 337 \kpcc. From the
best fit model (cstat/dof=284/262, see Fig.\ref{fig:spectrum}) the
unabsorbed flux in the 0.1-2.4 keV band is $2.01\pm0.04 \times
10^{-12}$ \fxunits, which corresponds to an unabsorbed luminosity of
$4.84\pm0.09 \times 10^{43}$ \lxunits\ in the rest frame band 0.1-2.4
keV.  The quoted errors on flux and luminosity were obtained by XSPEC
using a Montecarlo procedure.
Assuming that the cluster emission profile follows the best fit
$\beta$-model derived above, the luminosity within $R_{500}=1100$ \kpcc, 
as estimated from the Arnaud et al. (\citeyear{arn05})
scaling relation, is $L_{\rm X,500} = 1.57\pm0.14 \times 10^{44}$
\lxunits\ in the 0.1-2.4 keV rest frame band.  Errors in the
luminosity were determined including both the spectral errors and the
uncertainties in the $\beta$-model parameters. The aperture of
337 \kpc used as reference contains 31\% of the total luminosity
estimated within $R_{500}$.  The revised upward \chandra\ luminosity
makes A523 more consistent with the envelope of the non-cool core
clusters in the $L_{\rm X}$-$T$ relation as shown in Fig.\ref{fig:lt}.

\section{ANALYSIS OF RADIO DATA}
\label{radio}

\subsection{Radio images}
\label{radiodata}

We re-analysed archival VLA data at 1.4 GHz in
C-array and D-array configuration (AB1180 and AR690 projects). These
data had been published in G11 and van Weeren et
al. (\citeyear{vanwee11a}). For this study, we also considered the
polarization information available from the AR690 project.  The data
were reduced following standard procedures using the NRAO's
Astronomical Image Processing System (AIPS) package.

\begin{figure}
\centering
\resizebox{\hsize}{!}{\includegraphics{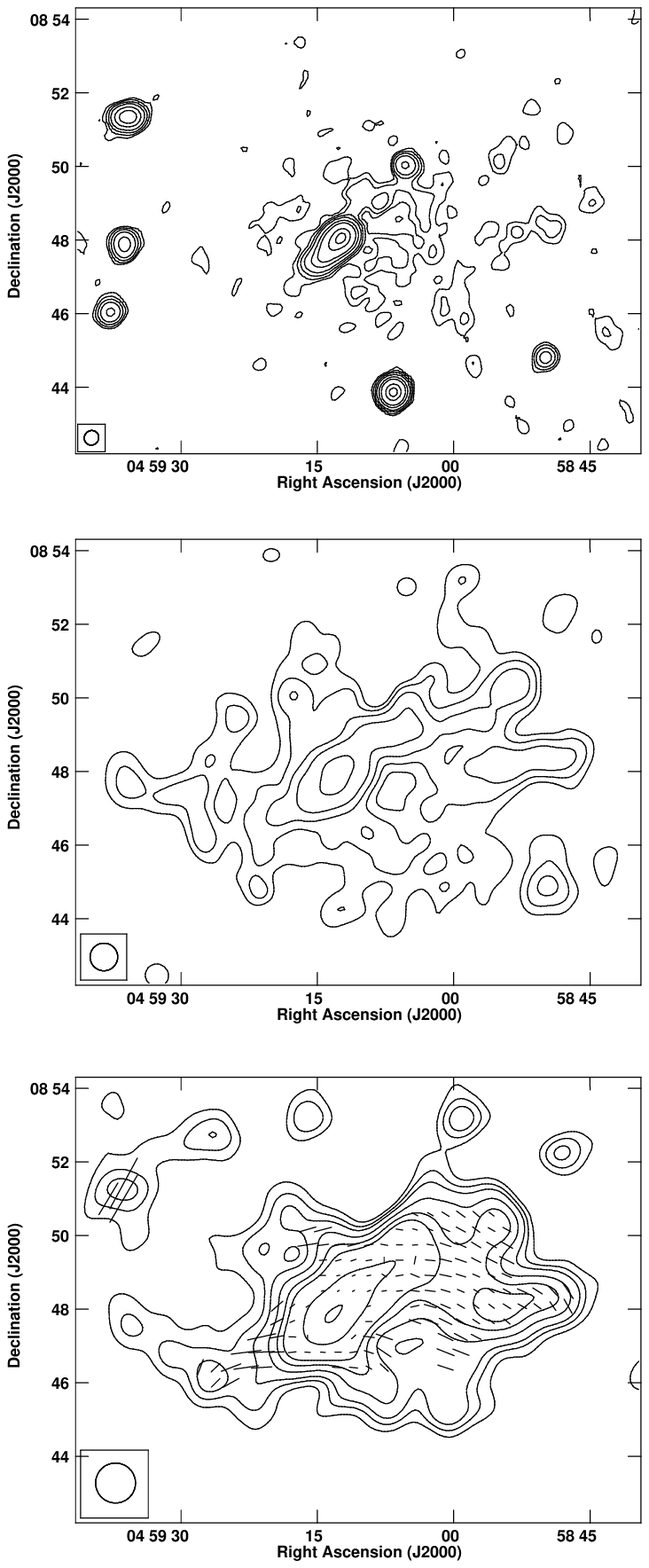}}
\caption{Top panel: total intensity contours at 1.4 GHz with a
  FWHM of 24.4\arcss$\times$23.2\arcs (PA=$-42\degree$). The contour
  levels are drawn at: 0.025, 0.05, 0.1, 0.2, 0.5, 1, 3, 5 mJy/beam.
  The sensitivity (1$\sigma$) is 0.012 mJy/beam. Middle panel: total
  intensity contours at 1.4 GHz with a FWHM of
  45\arcss$\times$45\arcss.  The contour levels are drawn at: 0.2,
  0.5, 1, 2, 5, 10 mJy/beam.  The sensitivity (1$\sigma$) is 0.07
  mJy/beam.  Bottom panel: total intensity contours and polarization
  vectors at 1.4 GHz with a FWHM of 65\arcss$\times$65\arcss.  The
  contour levels are drawn at: 0.2, 0.4, 0.6, 1, 1.5, 2, 4, 6
  mJy/beam.  The sensitivity (1$\sigma$) is 0.1 mJy/beam.  The lines
  give the orientation of the electric vector position angle (E-field)
  and are proportional in length to the FPOL
  (1\arcss$\simeq$1.25\%).  In all the panels, the FOW of the
  image is $\sim 1.8 \times 1.4$ \hh.}
\label{radioimage}
\end{figure}

In Fig.~\ref{radioimage} (top panel) we present an image at a
relatively high resolution (24.4\arcs $\times$ 23.2\arcss,
PA=$-42\degree$), where the central head--tail radio galaxy is clearly
visible as well as the central region of the diffuse emission. Other
point-like sources are present. The image noise level is 0.012
mJy/beam.  From this image it is clear that no connection is present
between the radio tail and the diffuse emission (cfr. with
Fig.~\ref{radioimage} -- middle panel).

In Fig.~\ref{radioimage} (middle panel) we present an image at a
resolution of 45\arcss$\times$45\arcs with the discrete sources
subtracted. The image noise level is 0.07 mJy/beam.  The discrete
sources were subtracted in the ($u,v$) plane by using the AIPS task
UVSUB.  This image confirms the presence of a diffuse low-surface
brightness radio emission at the cluster center.  The total flux
density at 1.4 GHz is 72$\pm$3 mJy.  This flux density value
corresponds to a radio power of $P_{\rm 1.4\,GHz} =
1.98\times10^{24}$\,{\rm W~Hz}$^{\rm -1}$.  The quoted error on the
flux density not only is due to the noise but also includes uncertainties
in the source subtraction and in the source size.  The source angular
size is $\simeq$11\arcmm$\times$8\arcmm.

In Fig.~\ref{radioimage} (bottom panel), we present a total intensity
image and polarization vectors at a resolution of
65\arcss$\times$65\arcs with the discrete sources subtracted. The
image noise level is 0.1 mJy/beam.  The vectors give the orientation
of the electric vector position angle (E-field) and are proportional
in length to the fractional polarization (FPOL).  In the FPOL
image we considered as valid those pixels where the
FPOL was above $3\sigma_{\mathrm FPOL}$.

From a detailed analysis of the radio morphology of the diffuse
emission, we can derive the following conclusions.  i) There is no
connection among discrete sources and the diffuse source. In
particular, the head--tail galaxy located close to the center of the
extended source is elongated on the opposite side with respect to the
brightest region of the diffuse source (see Fig.~\ref{radioimage}, top
panel).  ii) The residuals from the subtraction of discrete sources
are at a very low level and do not affect the diffuse source
properties (compare Fig.~\ref{radioimage}, top and middle panels).
iii) The diffuse source is generally elongated along the ESE-WNW
direction. In addition, a SSW-NNE elongation is present.  The radio
emission shows an irregular shape, with a filamentary structure
characterized by a bright ridge close to the center.  However, there
is no evidence of a transverse flux asymmetry as usually found in
relic sources due to the propagation of a shock wave (see Feretti et
al. \citeyear{fer12}).  Moreover, as clearly seen in
Fig.~\ref{radioimage} (top panel), this bright ridge is completely
resolved increasing the angular resolution and no compact feature is
present, contrary to most of the classic relics found so far (see
Feretti et al. \citeyear{fer12} and references therein).

Therefore, in the following section we assume that the diffuse radio
source in A523 is a radio halo. More specifically, we suppose that the
synchrotron plasma permeates the entire cluster volume rather than
being confined to a thin-layer in the peripheral region as in the case
of a radio relic.  Under this assumption and considering the polarization
associated with the radio halo, we analysed the properties
of the total intensity and polarized emission with the help of
numerical simulations performed with the software FARADAY (Murgia et
al. \citeyear{mur04}), and we compared the results with the properties of the
other radio haloes known in the literature.

\subsection{Radio brightness profile}
\label{radioprof}

\begin{figure*}
\centering
\includegraphics[width=18cm]{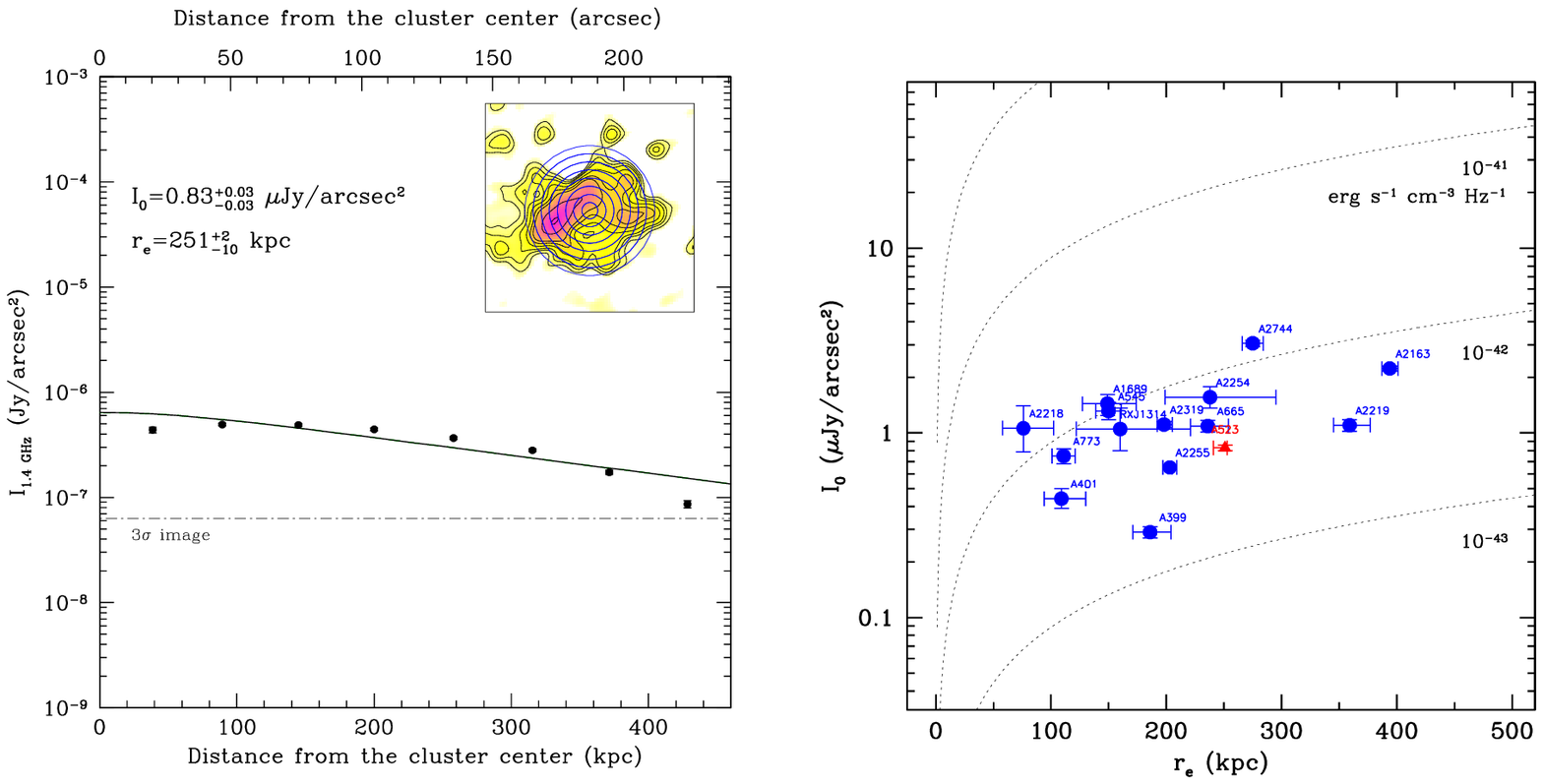}
\caption{ {\em Left panel}: the azimuthally averaged brightness profile of
  the radio halo emission in A523. The profile has been calculated in
  concentric annuli centred on the radio centroid, as shown in the
  inset panel.  The horizontal dashed-dotted line indicates the
  3$\sigma$ noise level of the radio image, while the continuous line
  indicates the best-fit profile described by an exponential law.
  Right panel: best-fit central brightness $I_{\rm 0}$ at 1.4\,{\rm GHz}
  versus the length scale $r_{\rm e}$ of A523 in comparison with
  azimuthally averaged brightness profiles of radio haloes taken from
  the literature (Murgia et al. \citeyear{mur09}, Murgia et
  al. \citeyear{mur10}, Vacca et al. \citeyear{vac11}, and references
  therein).  The dotted lines indicate regions of constant synchrotron
  emissivity. }
\label{profilo}
\end{figure*}

In the left panel of Fig.~\ref{profilo} we show the azimuthally averaged 
radio halo brightness profile of A523 obtained from the 
lower resolution (FWHM=65\arcss$\times$65\arcss) image with the discrete 
sources subtracted. Each data point represents the average brightness in 
concentric annuli of 30\arcs (about half beam width),
centred on the radio centroid.

We modeled the radio brightness profile, $I(r)$, with an exponential
of the form $I(r)=I_{0}e^{-r/r_{e}}$, whose best-fit is shown in the
left panel of Fig.~\ref{profilo}.  The proposed method for deriving the
radio brightness and the length scale of diffuse emission (Murgia et
al. \citeyear{mur09}), is relatively independent of the sensitivity of the
radio observation. The exponential model is attractive in its
simplicity and involves a minimal set of free parameters but,
obviously, it cannot account for the local deviations from the
circular symmetry of the diffuse emission.  The best-fit of the
exponential model yields a central brightness of $I_{0}$=0.83
$\mu$Jy/\arcs$^2$ and a length scale $r_{e}$=$251\pm10$ \kpcc.

In the right panel of Fig.~\ref{profilo}, we show $I_{\rm 0}$ versus
$r_{\rm e}$ of A523 in comparison with a set of radio haloes analysed
in the literature (Murgia et al. \citeyear{mur09}, Murgia et
al. \citeyear{mur10}, Vacca et al. \citeyear{vac11}).  As previously pointed
out, radio haloes can have quite different length scales, but their
emissivity is remarkably similar from one halo to another.  A523
populates the same region of the $I_{\rm0}-r_{\rm e}$ plane as the
other radio haloes.  Therefore, the diffuse total intensity emission of
A523 seems in good agreement with the extrapolation of the properties
of the other radio haloes known in the literature.

\begin{figure*}
\centering
\includegraphics[width=10cm]{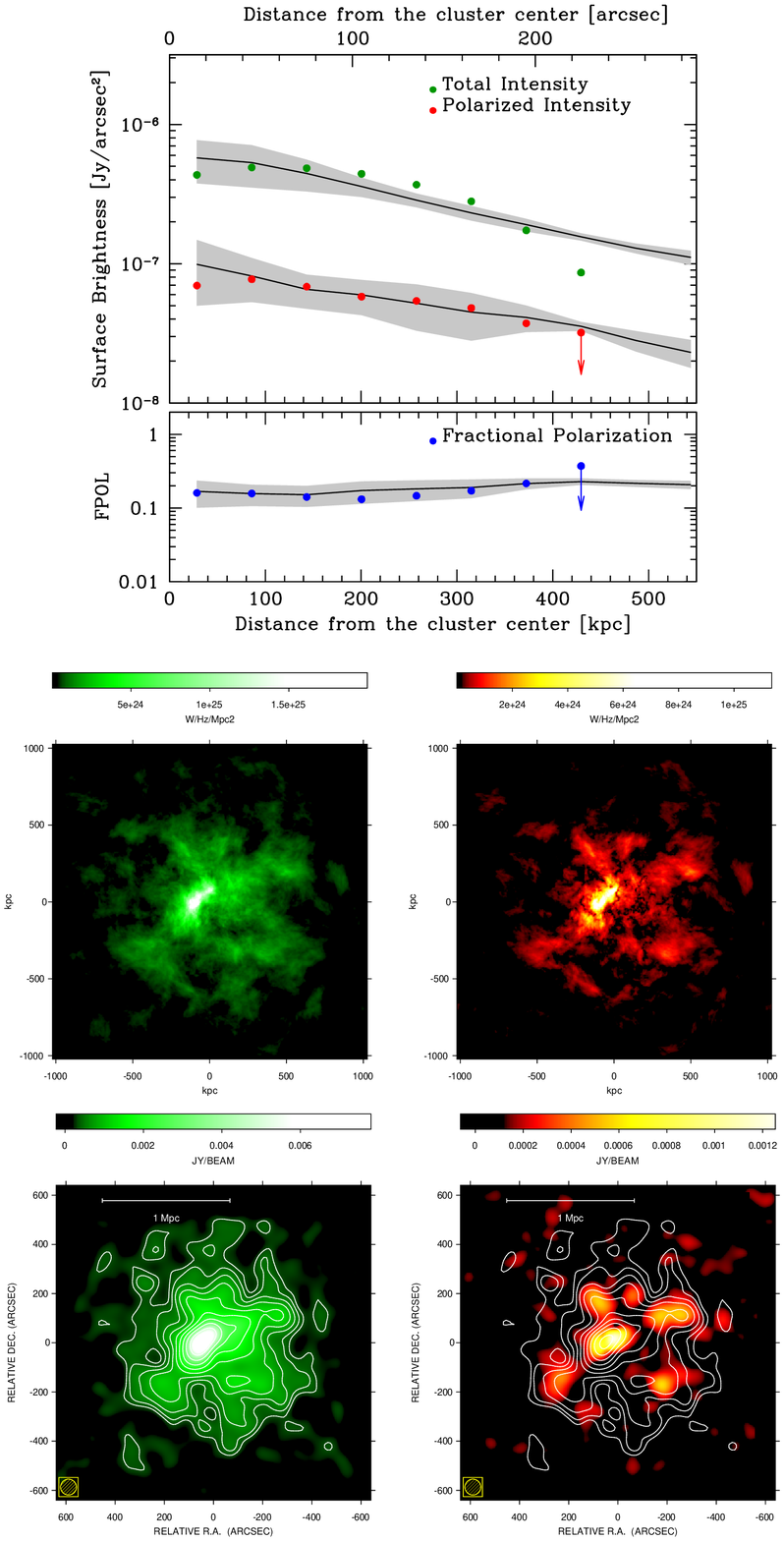}
\caption{Top panels: azimuthally averaged radio-halo brightness
  profiles of the total intensity I (green dots), polarized intensity
  $P$ (red dots), and FPOL (blue dots).  The
  profiles have been calculated in concentric annuli, as shown in
  Fig.~\ref{profilo}.  The continuum line and the shade region in the
  top panels describe the mean and the sigma of a set of simulations
  with different random seeds performed with a magnetic field model
  which is able to describe the data quite well. This model is a
  Kolmogorov magnetic field power spectrum with a central magnetic
  field strength of 0.5 $\mu$G which fluctuates in a range of scales
  from a few \kpc up to $\sim 1$ \h and decreases with the gas density
  as B(r) $\propto$ n$_e$(r)$^{0.5}$.  Middle panels: example of
  a simulated radio halo at full resolution obtained with the above
  magnetic field model. Left and right panels refer to the total
  intensity and polarized surface brightness images,
  respectively. Bottom panels: the images show as the simulated
  radio halo would appear when observed at the same resolution and
  sensitivity of A523.  The white contour levels refer to the total
  intensity image at 65\arcs resolution. Contour levels start at 0.3
  mJy/beam and increase by a factor of $\sqrt2$.  }
\label{profilo_simulato}
\end{figure*}

Morphological similarities between radio and X-ray images have been
found in a number of clusters hosting a radio halo (Govoni et
al. \citeyear{gov01a}) and one feature of A523 is that the
point-to-point radio and X-ray emission appears quite different (see
also Sect.~\ref{intro} and refs. therein). However, other irregular
and asymmetric haloes have been found in the literature (Govoni et
al. \citeyear{gov12}).  Feretti et al. (\citeyear{fer12}) investigated
the statistics of the offset between the peak of the radio halo and of
the X-ray emission of the cluster.  They have found that the offsets
can reach several kpc and they become more relevant for haloes of
smaller size.  A possible explanation for this behaviour can be
attributed to the cluster magnetic field power spectrum.  Indeed,
magnetic field simulations (Murgia et al. \citeyear{mur04}, Govoni et
al.  \citeyear{gov06}, Vacca et al. \citeyear{vac10}) show that
clusters having most of the magnetic field energy on large spatial
scales may produce radio haloes characterized by comparatively high
polarization emission with a filamentary structure. Instead,
unpolarized radio haloes with a regular morphology are expected in
clusters in which most of the magnetic field energy is on small
spatial scales.  In agreement with the simulation expectations
filaments of polarized emission associated with a radio halo have been
detected in A2255 (Govoni et al. \citeyear{gov05}) and MACS
J0717.5+3745 (Bonafede et al. \citeyear{bon09}).

In addition to the radio brightness profile of the total intensity,
in Fig.~\ref{profilo_simulato} we show the observed
brightness profiles of the polarized intensity and FPOL
calculated in concentric annuli, as shown in Fig.~\ref{profilo}.
The observed profiles are traced up to a projected 
distance from the cluster center of $\simeq$450 \kpcc.
The FPOL profile is rather constant 
at $\simeq$15-20\%.

To investigate if the observed FPOL levels and the
distorted structure of A523 can be observed in a radio halo, we
simulated 3D magnetic fields with a single power-law power spectrum of
the magnetic field fluctuations $|B_k|^2\propto k^{-n}$ with a
Kolmogorov spectral index $n=11/3$.  We generated synthetic radio halo
images by illuminating 3D magnetic field models with a population of
relativistic electrons. At each point, on the computational grid, we
calculated the total intensity and the intrinsic linear polarization
emissivity at 1.4 GHz, by convolving the emission spectrum of a single
relativistic electron with the particle energy distribution of an
isotropic population of relativistic electrons.  The polarization
images were obtained by taking into account that the polarization
plane of the radio signal is subject to the Faraday rotation as it
traverses the magnetized ICM. We modelled the gas
density of A523 with the $\beta$-model parameters $r_c$=181 \kpc
$\beta$=0.42, and a central gas density of n$_0$=$1.1\times10^{-3}$
cm$^{-3}$ (see Sect.~\ref{ximage}).
 
The results of the simulations are presented in
Fig.~\ref{profilo_simulato}. By adopting a Kolmogorov spectral index
for the magnetic field fluctuations and by considering a magnetic
field which decreases with the gas density as B(r)$\propto$
n$_e$(r)$^{0.5}$, the FPOL levels can be well reproduced by a magnetic
field with a central strength $B_0 \simeq$ 0.5 $\mu$G which fluctuates
in a range of scales from a few \kpc up to $\sim 1$ \h.  The continuum
line and the shade region describe the mean and the sigma of a set of
simulations (with different random seeds) performed with the magnetic
field model described above.  The total intensity depends on both the
magnetic field and on the population of relativistic electrons, while
the FPOL mostly depends on the cluster magnetic field
correlation-length and strength.  Once constrained the magnetic field
on the basis of the FPOL profile, it is possible to investigate the
population of relativistic electrons.  By assuming the equipartition
between magnetic field and relativistic electrons, the profiles can be
well described by assuming a population of relativistic electrons with
an electron energy spectral index $\delta=3$, and a low and high
energy cut-offs of the energy spectrum of $\gamma_{\rm min}$=3500 and
$\gamma_{\rm max}$=$1.5\times10^4$, respectively.

In the middle panels of Fig.~\ref{profilo_simulato} we show an example
of a full resolution total intensity (left) and polarized intensity
(right) image obtained with the above magnetic field model. In the
bottom panels of Fig.~\ref{profilo_simulato} we show as the simulated
radio halo would appear when observed at the same resolution 65\arcs
and sensitivity (0.1 mJy/beam) of A523.

Therefore we can conclude that a magnetic field model with a central
strength $B_0 \simeq$0.5 $\mu$G which fluctuates over a large spatial
scale is able to explain the presence of a polarized radio halo at a
level of 15-20\% characterized by a distorted radio morphology with a
significant displacement from the X-ray gas distribution.  In the
simulation shown in Fig.~\ref{profilo_simulato} the radio/X-ray
displacement is $\simeq$ 80 \kpcc.

\section{SUMMARY OF RESULTS AND DISCUSSION}
\label{disc}

We present new insights into the structure of the A523 galaxy cluster
from our multiwavelength analysis. The main results obtained from
optical, X-ray, and radio data are summarized in the following and
then discussed to derive a scenario consistent with the available
observational picture.

\subsection{Results from our multiwavelength analysis}
\label{dres}

On the basis of 80 member galaxies, we present the first measure of
the mean cluster redshift $\left<z\right>=0.1040\pm0.0004$ --
previously only $z_{\rm BCG1}$ was available -- and of the LOS
velocity dispersion of the galaxy population $\sigma_{\rm
V}=949_{-60}^{+80}$
\kss.  Our analysis of both spectroscopic and photometric optical data
confirms the bimodal structure of the cluster elongated in the SSW-NNE
direction, as already suggested by the analysis of old plate images
(G11), with the two subclusters $\sim 0.75$ \h apart in the plane of
sky.  We confirm that the northern subcluster is dominated by BCG1, the
brightest cluster galaxy, and identify BCG2 as the dominant
galaxy that lies in the southern cluster. The redshift data allow us to
detect that the two subclusters are (little) separated in the velocity
space and we estimate the relative LOS velocity $\Delta_{V,{\rm
    rf}}=100$-650 \ks in the cluster rest frame. Suggestions for a
more complex structure come from the analysis of the
spectroscopic and photometric samples, in particular for one or two
clumps tracing the NW-SE (or NWN-ESE) direction (see
Figs.~\ref{figk2z}, \ref{figdedica3D} and \ref{figk2}).

The analysis of \chandra\ data allows us to support with quantitative
measures the visual indication of a disturbed cluster already reported
by G11 on the basis of \rosat\ data. The hot ICM structure is now much
better defined.  The X-ray surface brightness is quite elongated
towards NNE, but there is no longer evidence of bimodality (cf. our
Fig.~\ref{fig:imax} and Fig.~1 of G11).  The peak of the surface
brightness is clearly offset with respect to both the two BCGs,
$\siml $ 0.5 and $\siml $ 0.2 \h from BCG1 and BCG2, respectively.  We
present the first measure of the global X-ray temperature, $kT_{\rm
  OUT}=5.3\pm0.3$ keV, and find no evidence for a cool core. Very
interestingly, the $kT$ map shows evidence for a higher temperature in
the northern region. Our new measure of the X-ray luminosity, $L_{{\rm
    X},500} = 1.6 \times 10^{44}$ \lxunits\ in the 0.1-2.4 keV rest
frame band, makes A523 more consistent with the observed $L_{\rm
  X}$-$T$ relation with respect to the previous \rosat\ estimate,
although still lying in the low luminosity regions of the envelope
(see Fig.~\ref{fig:lt}).

Our analysis of VLA data confirms the presence of an extended, diffuse
radio source having a maximum projected size of ${\rm LLS} \sim 1.3$
\hh. The revised estimate of the radio power, $P_{\rm 1.4\,GHz} =
2.0\pm 0.1 \times10^{24}$\,{\rm W~Hz}$^{\rm -1}$, is slightly higher
than the G11 estimate. Very interestingly, in addition to the main
ESE-WNW elongation, our new analysis shows that the radio source has a
minor SSW-NNE elongation almost aligned with the main optical/X-ray
cluster elongation (see Fig.~\ref{radioimage}, middle panel, and
Fig.~\ref{figimage}).  The radio source permeates the region between
the two subclusters and is classified as a radio halo.  In comparison
to other clusters hosting radio haloes, A523 is highly peculiar in the
$P_{\rm 1.4\,GHz}$-$L_{\rm X}$ plane having a higher radio power (or
lower X--ray luminosity) than expected (see also Sect.~\ref{dsca}).
Despite this, A523 is typical among clusters hosting radio haloes
since it populates the same region of the $I_0$-$r_e$ plane (see
Fig.~\ref{profilo}, right panel) and the $P_{\rm 1.4\,GHz}$-LLS plane
(see e.g., Fig.~7 of Feretti et al. \citeyear{fer12}).  The radio
emission is clearly offset from the X-ray emission and the
radio/X-ray peaks offset is $\sim$ 0.3 \hh.  We also detect a modest
polarization (${\rm FPOL} \sim 15-20\%$), unusual in radio haloes
since, so far, a polarized signal has been detected only in a couple
of other radio haloes. Both the observed radio/X-ray offset and
polarization might be the result of having most magnetic field energy
on large spatial scales as we show using an ad hoc set of simulations
(see Sect.~\ref{radioprof}).

\subsection{Cluster mass estimate}
\label{dmas}

Both our optical and X-ray data indicate that A523 is a massive
cluster. Using the theoretical relation between mass and velocity
dispersion of Munari et al. (\citeyear{mun13}; Eq.~1 checked on
simulated clusters), the derived cluster mass is $M_{200,{\rm
opt}}(<R_{200,{\rm opt}}=1.9 \hhh)=9.0$ \mquaa, with related
uncertainties of 8\% and 23\% on $R_{200,{\rm opt}}$ and $M_{200,{\rm
opt}}$, as propagated from the error on $\sigma_{\rm V}$.  An
additional 10\% of uncertainty on mass is indicated by the scatter
around the theoretical relation which, however, does not take into
account the cluster asphericity in the velocity ellipsoid (e.g.,
Wojtak \citeyear{woj13}).  From our measure of $kT_{\rm OUT}$ and the
scaling relation of Arnaud et al. (\citeyear{arn05}), the derived
cluster mass is $M_{200,{\rm X}}(<R_{200,{\rm X}}\sim 1.8 \hhh)\sim
7$ \mqua in good agreement with the above optical estimate.

\subsection{A merger scenario}
\label{dmer}

According to the main optical and X-ray features (but see
Sect.~\ref{dsug}), A523 can be described as a binary head--on merger
(BHOM) after the primary collision. In fact, it is characterized by
two important galaxy subclusters and typical X-ray features noted
in simulations (e.g., Ricker \& Sarazin
\citeyear{ric01}; Poole et al. \citeyear{poo06}). It also meets the typical
BHOM selection criteria as listed by Mann \& Ebeling
(\citeyear{man11}) that are the non-concentric X-ray contours and the
large offset of the two BCGs from the X-ray peak.

The alignment of the directions defined by the two subclusters and by
the elongation of X-ray isophotes indicate that SSW-NNE is the
direction of the merging axis.  The direction of the elongation of the
large, diffuse BCG1 halo agrees, too.  The small LOS velocity
difference between the two subclusters indicates that the merger axis
is almost perpendicular to the LOS or, alternatively, that the two
subclusters are close to the turn around point. The first hypothesis
is quite more reliable when analysing the merger kinematics with the
use of the simple analytical bimodal model (e.g., Girardi et
al. \citeyear{gir08} and refs. therein). According to our
observational results, the relevant parameters of the two--body model
are $V_{\rm rf}=100$-650 \ks for the relative LOS velocity,
$D=0.75$ \h for the projected distance, and $M_{\rm
system}=0.7$-0.9 \mqua for the mass of the system.  A reliable
assumption for the time relative to the core crossing is $t=0.1$-0.3
Gyr since a few $10^8$ years is the radiative lifetime of
relativistic electrons losing energy. Moreover, comparable values have
been found in previous studies of merging clusters hosting radio
haloes, e.g. $t=0.1$-0.2 Gyr is found in the case of the Bullet cluster
(Markevitch et al. \citeyear{mar02}), and $t=0.2$-0.3 Gyr in Abell 520
(Girardi et al. \citeyear{gir08}). Outgoing solutions are the only
acceptable ones and $\alpha$, the projection angle between the merging
axis and the plane of the sky, is at most $\sim 10$\degree\ in the
$V_{\rm rf}=100$ \ks case and at most $\sim 30$\degree\ in the $V_{\rm
rf}=650$ \ks case. As for the mass ratio, both our spectroscopic and
photometric optical data indicate that the southern cluster is
comparable/slightly richer than the northern cluster, suggesting that
the mass ratio is less than 2:1 (cfr. the relative densities and
galaxy contents in Tables~\ref{tabdedica2dz} and \ref{tabdedica2d}).

Evidence of a merger in A523 provides additional support to the idea
of a strong connection between the presence of a radio halo and an
active dynamical status of the host cluster (see Sect.~\ref{intro} and
refs. therein). In particular, A523 also agrees with other halo radio
clusters in the relations between the X-ray indicators of
substructure, i.e.  $c$ versus $w$, $w$ versus $P_3/P_0$, $c$ versus $P_3/P_0$
(Cassano et al. \citeyear{cas10}). Moreover, the BHOM scenario rules
out the possibility that the radio source is a relic instead than a
halo. In fact, in the outgoing shock-driven relic hypothesis we would
expect that the subcluster, slowed down by gravitational effects, is
preceded by the relic in its outgoing motion as shown by numerical
simulations (e.g., Springel \& Farral \citeyear{spr07}; Mastropietro
\& Burkert \citeyear{mas08}), while the radio source in A523
is centred between the centres of the two subclusters.

\subsection{Scaling relations}
\label{dsca}

An interesting global cluster parameter is the value of the ratio
between the energy per unit mass of galaxies to that of ICM as
parametrized with $\beta_{\rm spec}=\sigma_V^2/(kT/\mu m_{\rm p})$,
where $\mu=0.58$ is the mean molecular weight and $m_{\rm p}$ the
proton mass. The value $\beta_{\rm spec}=1$ indicates the
density-energy equipartition between ICM and galaxies.  The mean
$\beta_{\rm spec}$ value observed for massive clusters is consistent
with unity both in nearby (Girardi et
al. \citeyear{gir96}, \citeyear{gir98}) and in distant systems out to
$z\sim 0.4$ (Mushotzky \& Scharf \citeyear{mus97}).
Fig.~\ref{figprof} (bottom panel) shows how good is the agreement
between X-ray temperature and velocity dispersion in the case of A523.

Numerical simulations show as both X-ray temperature and velocity
dispersion rise due to the cluster merger, e.g., the temperature
peaks either during the core-crossing or just after and then declines
(Ricker \& Sarazin \citeyear{ric01}; Mastropietro \& Burkert
\citeyear{mas08}), and similarly the velocity dispersion.  However, a
small enhancement of velocity dispersion is expected in the case of a
merger axis perpendicular to the LOS ($\sim$20\% according to Pinkney
et al.  \citeyear{pin96}) and A523 is indeed quite normal in the other
relation involving global optical and X-ray properties.  In fact, when
considering our estimate of the bolometric X-ray luminosity ($L_{{\rm
X,bol},500} = 3.44 \times 10^{44}$ \lxunits) and of the velocity
dispersion ($\sigma_V=950$ \kss), A523 lies very close to the $L_{\rm
X}$-$\sigma$ relation as fitted by Zhang et
al. (\citeyear{zha11}). Moreover, the A523 values are still acceptable
in the $L_{\rm X}$-$T$ plane (as discussed in Sect.~\ref{lt}).  We
conclude that the agreement between $\sigma_V$ and $T$ quantities and
their use as mass proxies are likely reliable.

As in the case of A523, most clusters showing radio haloes have a large
gravitational mass, e.g., larger than $0.7\times 10^{14}$ M$_{\sun}$
within $2$ Mpc (Giovannini \& Feretti \citeyear{gio02}; see also
clusters analysed in our DARC program, Girardi et al. \citeyear{gir11}
and refs. therein). In particular, when considering our estimate of
the mass of A523 ($M_{500}\sim 5$-6 \mquaa, $M_{500}\sim 1/1.4 \times
M_{200}$ for a typical NFW profile), this cluster is consistent with
the relation between mass and radio power as traced for clusters
hosting radio haloes (see fig.~3 of Cassano et al. \citeyear{cas13}).

\begin{figure}
\centering 
\includegraphics[width=8cm]{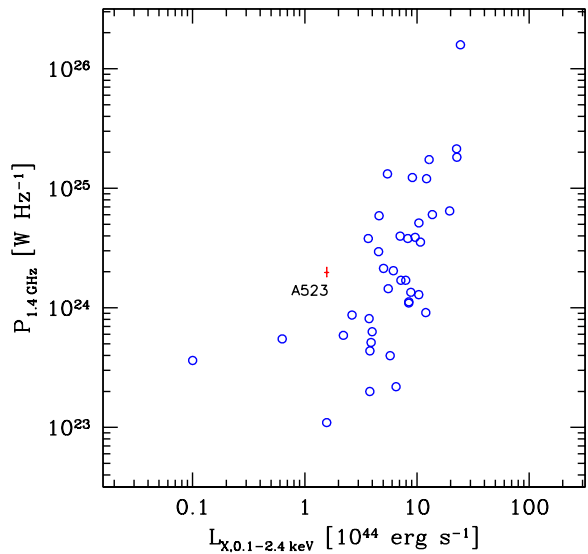}
\caption{Total halo radio power at 1.4 GHz versus cluster X-ray luminosity
in the 0.1-2.4 keV band. Blue circles refer to clusters plotted by
Feretti et al. (\citeyear{fer12}). The position of A523 is indicated
by the red cross and is based on our study.  }
\label{figPrLx}
\end{figure}

In spite of our revised upward X-ray luminosity, the peculiarity of
A523 in the $P_{\rm 1.4\,GHz}$-$L_{\rm X}$ plane, already claimed by
G11, is not solved and, rather, it should be considered more reliable
due to the improvement of the data. In Fig.~\ref{figPrLx} A523 is
compared to other $\sim 40$ clusters as taken from Feretti et
al. (\citeyear{fer12}).  According to the relation fitted on $\sim 25$
clusters by Cassano et al. (\citeyear{cas13}, see their fig.~2 and
table~3), A523 is under luminous in X-ray by a factor of $\simg$ 4 or
over luminous in radio by a factor of $\simg$ 24, which is inconsistent
with the small uncertainties associated with the observational
measures for A523 or with the scatter around the relation.

\subsection{Suggestions for  a more complex scenario}
\label{dsug}

The peculiarity of A523 is not limited to the $P_{\rm
  1.4\,GHz}$-$L_{\rm X}$ relation. Although our simulations in
  Sect.~\ref{radioprof} show that a radio/X-ray peak offset can be
  explained in the context of a magnetic field energy on large spatial
  scales and suggest that peculiar halo elongations can be observed
  (see Fig.~\ref{profilo_simulato}, middle and bottom left panels), it
  is still not obvious that we can explain a case like A523, where the
  halo elongation is perpendicular to the merging axis as traced by
  X-ray and optical data.

Looking for a possible explanation of the A523 peculiarity, we note
that, although earlier we discuss the A523 phenomenology in the
framework of a BHOM scenario, it is known that distinguishing between
binary and complex mergers is not easy (e.g., see the relevant
discussion in Mann \& Ebeling \citeyear{man11}).  In the case of A523,
both optical and radio data show some evidence in favour of a complex
merger suggesting a scenario where A523 is forming at the cross of two
filaments, along the SSW-NNE and ESE-WNW directions.  From the X-ray
side, a possible evidence in favour of a more complex dynamical status
comes from the enhanced temperature in the northern region,
more related to the radio halo position than the southern one.

In the context of a more complex scenario, the different main
directions traced by radio and X-ray+optical data might refer to the
two merging directions and, in particular, to be connected with the
different time-scales intervening, with the radio-halo better tracing
the most recent accretion phenomena, although this accretion does not
involve a lot of mass.  Unfortunately, also due to the contamination
of the background structure at $z\sim 0.14$, we need a larger redshift
sample to probe the ESE-WNW direction of accretion.  We also note that
our current redshift catalogue samples only the cluster region within
half $R_{200}$, making the cluster accretion phenomena only partially
traced.

\section*{Acknowledgments}

We thank the referee for useful and constructive comments.
M.G. acknowledges financial support from PRIN-MIUR 2010-11
(J91J12000450001).  V.V. is supported by the DFG Forschengruppe 1254
``Magnetisation of Interstellar and Intergalactic Media: The Prospects
of Low-Frequency Radio Observations''. This research was partially
supported by PRIN-INAF2014. This publication is based on observations
made on the island of La Palma with the Italian Telescopio Nazionale
{\em Galileo} (TNG) and the {\em Isaac Newton} Telescope, The
TNG is operated by the Fundaci\'on Galileo Galilei -- INAF (Istituto
Nazionale di Astrofisica) The INT is operated by the {\em Isaac
Newton} Group. Both telescopes are located in the Spanish Observatorio
of the Roque de Los Muchachos of the Instituto de Astrof\'isica de
Canarias (island of La Palma, Spain).  The scientific results reported
in this article are based in part on data obtained from the \chandra\
Data Archive.  The National Radio Astronomy Observatory is a facility
of the National Science Foundation operated under cooperative
agreement by Associated Universities, Inc.

%%%%%%%%%%%%%%%%%%%%%%%%%%%%%%%%%%%%%%%%%%%%%%%%%%



%\bibliographystyle{mnras}


% Don't change these lines
\bsp	% typesetting comment
\label{lastpage}
\end{document}

%% file: catalogA523a1.tex
%\documentclass[useAMS,usenatbib]{mn2e}
%\usepackage{graphicx}
%\newcommand {\ks} {km~s$^{-1} \;$}
%\newcommand {\kss} {km~s$^{-1}$}
%\newcommand {\mpc} {$Mpc \;$}
%\newcommand {\msun} {$h^{-1} \  M_{\odot} \;$}
%\newcommand {\m} {$M_{\odot} \;$}
%\newcommand {\ml} {$h \, M_{\odot}/L_{\odot} \;$}
%\newcommand {\mll} {$h \, M_{\odot}/L_{\odot}$}
%\newcommand{\vel}{\,{\rm km\,s^{-1}}}
%\newcommand{\tng}{\mathrm{T}}
%\newcommand{\sds}{\mathrm{S}}
%\newcommand{\tns}{\mathrm{T+S}}
%%
%\begin{document}

%\addtocounter{table}{-2}
\begin{table}
        \caption[]{Radial velocities of 132 galaxies in the field of
          A523. For each galaxy, the table lists right ascension
            and declination, $\alpha$ and $\delta$ (J2000);
            INT dereddened $r$-band magnitude, $r$; heliocentric
            radial velocity, $V$, with error, $\Delta V$.
          IDs in italics refer to nonmember galaxies. Galaxy ID~75 is
          the BCG1. Galaxy ID~56 is the BCG2.}
         \label{catalogA523}
              $$ 
        % \begin{array}{p{0.5\linewidth}l}
           \begin{array}{r c c r r}
            \hline
            \noalign{\smallskip}
            \hline
            \noalign{\smallskip}

\mathrm{ID} & \alpha,\delta\,(\mathrm{J}2000) & r\, &V\,& \Delta V\\
 &     (4^h,+8^o)             & &\mathrm{\,(\,km}&\mathrm{s^{-1}\,)}\\
            \hline
            \noalign{\smallskip}  

\textit{1}   &     58\ 47.73, 48\ 21.5 & 19.88 & 135555 & 112\\
\textit{2}   &     58\ 49.93, 47\ 54.6 & 18.26 &  42401 &  70\\
 3           &     58\ 52.20, 48\ 42.6 & 18.14 &  30189 &  51\\
 4           &     58\ 52.28, 55\ 44.5 & 17.86 &  32223 &  68\\
 5           &     58\ 53.38, 49\ 39.4 & 19.50 &  31578 &  59\\
 6           &     58\ 53.76, 46\ 54.4 & 19.07 &  30182 &  84\\
\textit{7}   &     58\ 54.06, 48\ 21.5 & 17.66 &  41879 &  44\\
\textit{8}   &     58\ 54.23, 48\ 42.9 & 15.73 &  41863 &  40\\
 9           &     58\ 54.74, 51\ 37.7 & 19.46 &  31600 &  73\\
10           &     58\ 54.78, 42\ 57.2 & 17.89 &  29600 &  55\\
11           &     58\ 54.97, 52\ 20.4 & 19.60 &  32078 & 132\\
\textit{12}  &     58\ 55.08, 55\ 44.5 & 18.99 &  42144 &  73\\
13           &     58\ 55.12, 43\ 54.2 & 18.29 &  32798 &  35\\
14           &     58\ 55.31, 50\ 16.4 & 17.82 &  29935 &  42\\
15           &     58\ 55.67, 51\ 08.5 & 18.40 &  32764 &  46\\
16           &     58\ 56.00, 50\ 27.2 & 17.69 &  31039 &  86\\
\textit{17}  &     58\ 56.53, 49\ 36.2 & 17.11 &  43442 &  35\\
\textit{18}  &     58\ 56.63, 54\ 33.2 & 18.46 &  42646 & 145\\
19           &     58\ 57.66, 48\ 49.7 & 17.49 &  31756 &  44\\
20           &     58\ 57.70, 41\ 21.6 & 18.20 &  29470 &  37\\
\textit{21}  &     58\ 57.90, 47\ 01.0 & 17.61 &  41165 &  46\\
\textit{22}  &     58\ 57.72, 45\ 41.8 & 18.86 &  63627 &  77\\
\textit{23}  &     58\ 57.76, 53\ 26.1 & 18.59 &  47438 &  88\\
\textit{24}  &     58\ 57.96, 53\ 06.8 & 19.31 &  47330 & 194\\
\textit{25}  &     58\ 58.67, 56\ 09.2 & 20.29 &  10848 & 101\\
26           &     58\ 58.91, 50\ 45.2 & 19.46 &  30398 &  70\\
27           &     58\ 58.93, 51\ 36.9 & 18.49 &  30549 &  59\\
\textit{28}  &     58\ 59.02, 45\ 10.2 & 17.18 &  23945 &  75\\
29           &     58\ 59.29, 43\ 57.7 & 18.68 &  31526 &  44\\
30           &     58\ 59.43, 47\ 47.7 & 17.90 &  32114 &  35\\
31           &     58\ 59.57, 45\ 19.6 & 19.42 &  31090 &  73\\
32           &     58\ 59.77, 44\ 49.6 & 18.23 &  30736 &  51\\
33           &     59\ 00.83, 47\ 09.5 & 17.94 &  32626 &  42\\
\textit{34}  &     59\ 00.90, 53\ 59.2 & 19.22 &  51820 & 101\\
35           &     59\ 01.15, 51\ 25.6 & 18.65 &  29595 &  66\\
\textit{36}  &     59\ 01.51, 46\ 45.5 & 17.66 &  40808 &  53\\
\textit{37}  &     59\ 01.90, 57\ 15.7 & 17.53 &  20563 &  73\\
38           &     59\ 02.12, 44\ 55.8 & 16.96 &  32114 &  35\\
39           &     59\ 02.73, 41\ 46.5 & 17.88 &  30866 &  44\\
40           &     59\ 02.99, 45\ 12.7 & 17.41 &  32133 &  44\\
41           &     59\ 03.28, 48\ 20.1 & 19.91 &  31782 & 103\\
\textit{42}  &     59\ 03.42, 53\ 12.3 & 19.26 &  41688 & 119\\
43           &     59\ 03.64, 42\ 33.8 & 18.23 &  31831 &  35\\

            \noalign{\smallskip}			 
            \hline					    
            \noalign{\smallskip}			    
            \hline					    
         \end{array}					 
     $$ 						 
         \end{table}					 
\addtocounter{table}{-1}				 
\begin{table}					 
          \caption[ ]{Continued.}
     $$ 
           \begin{array}{r c c r r}
            \hline
            \noalign{\smallskip}
            \hline
            \noalign{\smallskip}
\mathrm{ID} & \alpha,\delta\,(\mathrm{J}2000) & r\, &V\,& \Delta V\\
 &     (4^h,+8^o)     & &\mathrm{\,(\,km}&\mathrm{s^{-1}\,)}\\

            \hline
            \noalign{\smallskip}

\textit{44}  &     59\ 03.94, 56\ 32.7 & 19.99 & 109090 & 108\\
45           &     59\ 03.96, 47\ 15.4 & 17.74 &  30926 &  36\\
46           &     59\ 04.03, 41\ 46.7 & 16.52 &  30757 &  53\\
47           &     59\ 04.87, 45\ 17.8 & 18.49 &  29903 & 163\\
48           &     59\ 04.92, 46\ 56.0 & 19.22 &  31911 &  53\\
49           &     59\ 05.17, 43\ 13.5 & 20.50 &  31887 &  95\\
50           &     59\ 05.46, 52\ 25.7 & 18.93 &  31665 &  55\\
51           &     59\ 05.49, 42\ 41.0 & 18.28 &  31131 &  33\\
52           &     59\ 05.54, 45\ 55.1 & 17.64 &  30677 &  40\\
53           &     59\ 06.12, 45\ 21.6 & 16.51 &  30894 &  40\\
\textit{54}  &     59\ 06.23, 45\ 11.8 & 17.46 &  42229 &  62\\
55           &     59\ 06.49, 44\ 23.5 & 17.11 &  31001 &  40\\
\textbf{56}  &     59\ 06.59, 43\ 49.2 & 15.83 &  31222 &  44\\
\textit{57}  &     59\ 07.37, 48\ 54.5 & 18.40 &  41700 & 134\\
58           &     59\ 07.43, 49\ 20.7 & 19.00 &  31071 &  70\\
59           &     59\ 07.77, 41\ 05.2 & 18.25 &  32278 &  33\\
60           &     59\ 07.96, 50\ 49.5 & 18.28 &  31089 &  48\\
61           &     59\ 08.04, 50\ 13.3 & 18.00 &  30776 &  70\\
62           &     59\ 08.90, 45\ 43.8 & 19.32 &  32360 & 103\\
63           &     59\ 08.98, 43\ 51.1 & 18.55 &  31380 &  48\\
64           &     59\ 10.98, 51\ 04.0 & 19.71 &  30561 & 134\\
\textit{65}  &     59\ 11.03, 53\ 27.9 & 19.76 &  45429 & 100\\
66           &     59\ 11.17, 48\ 35.7 & 16.49 &  28498 &  68\\
67           &     59\ 11.18, 52\ 49.6 & 16.35 &  32901 &  44\\
68           &     59\ 11.32, 48\ 12.6 & 16.78 &  30143 &  40\\
\textit{69}  &     59\ 11.90, 54\ 49.1 & 18.43 &  60294 &  59\\
70           &     59\ 11.95, 44\ 14.0 & 18.45 &  31502 &  46\\
\textit{71}  &     59\ 12.11, 43\ 04.0 & 18.99 &  83329 &  57\\
72           &     59\ 12.65, 50\ 01.9 & 16.90 &  29941 &  42\\
73           &     59\ 12.86, 46\ 59.4 & 19.00 &  28631 & 114\\
\textit{74}  &     59\ 12.87, 53\ 32.8 & 18.23 &  41920 &  57\\
\textbf{75}  &     59\ 12.94, 49\ 41.1 & 14.68 &  31053 &  42\\
76           &     59\ 12.99, 49\ 14.2 & 17.98 &  30890 &  42\\
77           &     59\ 13.43, 47\ 15.6 & 17.28 &  30127 &  51\\
\textit{78}  &     59\ 13.91, 44\ 06.3 & 18.56 &  42410 & 110\\
79           &     59\ 14.12, 50\ 44.6 & 18.25 &  30403 &  44\\
80           &     59\ 14.39, 49\ 44.5 & 17.26 &  31107 &  64\\
\textit{81}  &     59\ 14.48, 43\ 32.3 & 19.48 &  50143 & 114\\
\textit{82}  &     59\ 14.49, 47\ 10.9 & 18.48 &  42176 &  66\\
83           &     59\ 14.73, 50\ 27.1 & 18.12 &  32894 &  42\\
\textit{84}  &     59\ 14.99, 55\ 57.6 & 20.01 &  59596 &  81\\
85           &     59\ 15.09, 51\ 52.1 & 18.81 &  30076 &  39\\
86           &     59\ 15.51, 48\ 02.2 & 17.40 &  30765 &  40\\

            \noalign{\smallskip}			    
            \hline					    
            \noalign{\smallskip}			    
            \hline					    
         \end{array}
     $$ 
         \end{table}
\addtocounter{table}{-1}
\begin{table}
          \caption[ ]{Continued.}
     $$ 
           \begin{array}{r c c r r}
            \hline
            \noalign{\smallskip}
            \hline
            \noalign{\smallskip}

\mathrm{ID} & \alpha,\delta\,(\mathrm{J}2000) & r\, &V\,& \Delta V\\
 &    (4^h,+8^o)    & &\mathrm{\,(\,km}&\mathrm{s^{-1}\,)}\\

            \hline
            \noalign{\smallskip}
   
87           &     59\ 16.00, 45\ 21.2 & 16.94 &  33072 &  99\\
88           &     59\ 16.18, 51\ 06.4 & 18.44 &  29114 &  62\\
89           &     59\ 16.38, 52\ 00.7 & 18.07 &  30772 &  33\\
\textit{90}  &     59\ 16.54, 43\ 03.2 & 18.69 & 108965 &  62\\
91           &     59\ 17.16, 52\ 58.3 & 19.20 &  30488 & 121\\
\textit{92}  &     59\ 17.62, 52\ 09.0 & 19.68 &  42064 &  77\\
93           &     59\ 18.10, 46\ 17.8 & 19.40 &  30659 &  44\\
94           &     59\ 18.16, 46\ 58.4 & 17.32 &  30767 &  35\\
\textit{95}  &     59\ 20.25, 45\ 01.1 & 16.03 &  20028 &  29\\
96           &     59\ 20.41, 54\ 46.8 & 16.69 &  31786 &  48\\
\textit{97}  &     59\ 20.54, 43\ 43.2 & 19.91 &  49204 & 119\\
98           &     59\ 20.99, 45\ 14.7 & 17.55 &  30772 &  57\\
99           &     59\ 21.77, 50\ 44.0 & 19.51 &  32911 & 130\\
100          &     59\ 22.56, 52\ 00.4 & 19.26 &  32790 & 123\\
101          &     59\ 23.26, 46\ 14.9 & 17.27 &  29947 &  37\\
102          &     59\ 23.32, 45\ 34.2 & 19.97 &  29535 &  86\\
103          &     59\ 23.32, 49\ 40.7 & 19.23 &  30121 &  92\\
104          &     59\ 25.05, 46\ 42.3 & 17.61 &  32083 &  59\\
\textit{105} &     59\ 26.67, 46\ 15.1 & 20.14 &  41988 & 198\\
106          &     59\ 27.19, 46\ 04.4 & 17.75 &  31814 &  51\\
107          &     59\ 29.68, 52\ 37.6 & 18.12 &  31245 &  95\\
\textit{108} &     59\ 30.46, 52\ 49.9 & 18.92 &  60220 & 103\\
109          &     59\ 30.60, 53\ 23.4 & 18.74 &  32014 & 136\\
\textit{110} &     59\ 32.03, 48\ 31.5 & 19.00 &  45191 & 100\\
111          &     59\ 32.60, 51\ 52.6 & 17.86 &  32103 &  88\\
\textit{112} &     59\ 33.12, 46\ 34.2 & 17.62 &  42361 &  44\\
\textit{113} &     59\ 33.33, 48\ 58.1 & 19.21 & 126550 & 128\\
\textit{114} &     59\ 33.97, 45\ 06.5 & 17.66 &  42559 &  59\\
115          &     59\ 34.46, 50\ 55.1 & 17.48 &  30722 &  48\\
116          &     59\ 35.19, 49\ 24.1 & 19.49 &  32440 &  92\\
\textit{117} &     59\ 36.06, 49\ 55.2 & 19.34 &  41935 &  53\\
\textit{118} &     59\ 36.06, 50\ 18.1 & 19.24 &  47985 & 112\\
\textit{119} &     59\ 36.70, 44\ 06.8 & 18.28 &  42416 &  90\\
\textit{120} &     59\ 36.84, 46\ 58.8 & 19.43 &  41824 & 123\\
121          &     59\ 38.24, 49\ 08.5 & 17.79 &  31662 &  44\\
\textit{122} &     59\ 38.70, 46\ 55.7 & 18.42 &  41001 &  40\\
\textit{123} &     59\ 38.79, 46\ 16.2 & 18.70 &  42861 &  81\\
\textit{124} &     59\ 39.18, 50\ 49.3 & 19.78 &  63094 & 106\\
\textit{125} &     59\ 40.78, 46\ 28.8 & 18.89 &  41288 &  57\\
\textit{126} &     59\ 42.75, 45\ 29.9 & 17.13 &  44699 &  40\\
\textit{127} &     59\ 43.55, 52\ 54.1 & 17.26 &  42060 &  31\\
\textit{128} &     59\ 43.84, 47\ 08.5 & 19.10 &  60116 & 103\\
\textit{129} &     59\ 45.06, 45\ 39.8 & 18.97 &  41287 &  99\\
\textit{130} &     59\ 48.43, 51\ 06.9 & 18.29 &  71136 & 134\\
131          &     59\ 48.83, 47\ 25.4 & 16.79 &  31365 &  46\\
\textit{132} &     59\ 56.68, 51\ 10.9 &   -.- &  75506 & 119\\
                                                     
            \noalign{\smallskip}			    
            \hline					    
            \noalign{\smallskip}			    
            \hline					    
         \end{array}
     $$ 
\end{table}

%\end{document}

%% file: tabdedica2dz.tex
\begin{table}
        \caption[]{Results of the 2D-DEDICA analysis from  the          spectroscopic sample.  For each subsample detected in the $z$ catalog, the table lists          the number of assigned member galaxies $N_{\rm S}$,          R.A. and Dec. of the density peak, the relative density with          respect to the highest peak $\rho_{\rm S}$, and $\chi^2$          value of the peak.}
         \label{tabdedica2dz}
            $$
         \begin{array}{l r c c r }
            \hline
            \noalign{\smallskip}
            \hline
            \noalign{\smallskip}
\mathrm{Subclump} & N_{\rm S} & \alpha({\rm J}2000),\,\delta({\rm J}2000)&\rho_{
\rm S}&\chi^2_{\rm S}\\
& & \mathrm{h:m:s,\degree:\arcmm:\arcs}&&\\
         \hline
         \noalign{\smallskip}
\mathrm{NNE(z)}& 40&04\ 59\ 12.9+08\ 50\ 08&1.00&16\\
\mathrm{SSW(z)}& 28&04\ 59\ 04.5+08\ 45\ 03&0.98&18\\
\mathrm{NW(z)} & 12&04\ 58\ 57.3+08\ 50\ 46&0.58& 8\\
              \noalign{\smallskip}
              \noalign{\smallskip}
            \hline
            \noalign{\smallskip}
            \hline
         \end{array}
$$
\end{table}

%% file: tabdedica2dback.tex
\begin{table}
        \caption[]{The 2D structure of the background structure based
          on spectroscopic data. For each subsample the table lists
          the same quantities as in Table~\ref{tabdedica2dz}.}
         \label{tabdedica2dback}
            $$
         \begin{array}{l r c c r }
            \hline
            \noalign{\smallskip}
            \hline
            \noalign{\smallskip}
\mathrm{Subclump} & N_{\rm S} & \alpha({\rm J}2000),\,\delta({\rm J}2000)&\rho_{
\rm S}&\chi^2_{\rm S}\\
& & \mathrm{h:m:s,\degree:\arcmm:\arcs}&&\\
         \hline
         \noalign{\smallskip}
\mathrm{ESE-Bgroup}& 11&04\ 59\ 37.9+08\ 46\ 30&1.00&6\\
\mathrm{NWN-Bgroup}& 10&04\ 58\ 55.1+08\ 48\ 17&0.58&4\\
              \noalign{\smallskip}
              \noalign{\smallskip}
            \hline
            \noalign{\smallskip}
            \hline
         \end{array}
$$
\end{table}

%% file: tabdedica2d.tex
\begin{table}
        \caption[]{Results of the 2D-DEDICA analysis from the INT photometric sample. For each subsample the table lists the same quantities as in Table~\ref{tabdedica2dz}.}
         \label{tabdedica2d}
            $$
         \begin{array}{l r c c r }
            \hline
            \noalign{\smallskip}
            \hline
            \noalign{\smallskip}
\mathrm{Subclump} & N_{\rm S} & \alpha({\rm J}2000),\,\delta({\rm J}2000)&\rho_{
\rm S}&\chi^2_{\rm S}\\
& & \mathrm{h:m:s,\degree:\arcmm:\arcs}&&\\
         \hline
         \noalign{\smallskip}
\mathrm{SSW(2D)}           & 35&04\ 59\ 07.6+08\ 44\ 20&1.00&20\\
\mathrm{NNE(2D)}           & 20&04\ 59\ 14.4+08\ 50\ 07&0.71&11\\
\mathrm{S-SSW(2D)}         & 40&04\ 59\ 03.4+08\ 40\ 58&0.52&12\\
\mathrm{WNW(2D)}           & 22&04\ 58\ 54.3+08\ 49\ 00&0.28&7\\
\mathrm{ESE(2D)}           & 16&04\ 59\ 38.3+08\ 45\ 49&0.23&7\\
              \noalign{\smallskip}
              \noalign{\smallskip}
            \hline
            \noalign{\smallskip}
            \hline
         \end{array}
$$
\end{table}

%% file: tabxmorpho.tex
\begin{table}

        \caption[]{Results of the morphological analysis of the X-ray
          image of A523. The table lists the power ratio $P_3/P_0$;
          the centroid shift $w$ as computed within a radius of (500
          h$_{70}^{-1}$ kpc); the concentration parameter $c$.  All
        the three indicators indicate that A523 is a disturbed cluster.
        }

         \label{tabxmorpho}
            $$
         \begin{array}{c c c}
            \hline
            \noalign{\smallskip}
            \hline
            \noalign{\smallskip}
\mathrm{P_3/P_0} &w                               & c \\
(\times 10^{-7}) & (500\,\mathrm{h_{70}^{-1}\,kpc})& \\
         \hline
         \noalign{\smallskip}
1.8^{+2.3}_{-1.6} & 0.025\pm0.002 & 0.095\pm0.003 \\
              \noalign{\smallskip}
             \noalign{\smallskip}
            \hline
            \noalign{\smallskip}
            \hline
         \end{array}
$$
         \end{table}